\documentclass[a4paper,fleqn,usenatbib]{mnras}
\usepackage{newtxtext,newtxmath}
\usepackage[T1]{fontenc}
\usepackage{ae,aecompl}
\usepackage{graphicx}
\usepackage{amsmath}
\usepackage{amssymb}
\usepackage{csquotes}

\title{Revealing the tidal scars of the Small Magellanic Cloud}

\author[De Leo et al.]{Michele De Leo$^{1}$\thanks{E-mail: m.deleo@surrey.ac.uk},
Ricardo Carrera$^{2}$, Noelia E. D. No\"el$^{1}$, Justin I. Read$^{1}$
\newauthor{Denis Erkal$^{1}$, Carme Gallart$^{3,4}$}
\\
$^{1}$Department of Physics, University of Surrey, Guildford, GU2 7XH, UK\\
$^{2}$INAF - Osservatorio Astronomico di Padova, Vicolo dell'Osservatorio 5, 35122 Padova, Italy\\
$^{3}$Instituto de Astrofisica de Canarias, Calle V\'{i}a L\'{a}ctea, La Laguna, E-38205, Tenerife, Spain\\
$^{4}$University of La Laguna. Avda. Astrof\'{i}sico Fco. S\'{a}nchez, La Laguna, E-38206, Tenerife, Spain\\
}

\date{Accepted XXX. Received YYY; in original form ZZZ}

\pubyear{2020}

\begin{document}
\label{firstpage}
\pagerange{\pageref{firstpage}--\pageref{lastpage}}
\maketitle

\begin{abstract}
\noindent 
Due to their close proximity, the Large and Small Magellanic Clouds (LMC/SMC) provide natural laboratories for understanding how galaxies form and evolve. With the goal of determining the structure and dynamical state of the SMC, we present  new spectroscopic data for $\sim$ 3000 SMC red giant branch stars observed using the AAOmega spectrograph at the Anglo-Australian Telescope. We complement our data with further spectroscopic measurements from previous studies that used the same instrumental configuration as well as proper motions from the \textit{Gaia} Data Release 2 catalogue. Analysing the photometric and stellar kinematic data, we find that the SMC centre of mass presents a conspicuous offset from the velocity centre of its associated $\mbox{H\,{\sc i}}$ gas, suggesting that the SMC gas is likely to be far from dynamical equilibrium. Furthermore, we find evidence that the SMC is currently undergoing tidal disruption by the LMC within 2\,kpc of the centre of the SMC, and possibly all the way into the very core. This is revealed by a net outward motion of stars from the SMC centre along the direction towards the LMC and an apparent tangential anisotropy at all radii. The latter is expected if the SMC is undergoing significant tidal stripping, as we demonstrate using a suite of $N$-body simulations of the SMC/LMC system disrupting around the Milky Way. Our results suggest that dynamical models for the SMC that assume a steady state will need to be revisited.
\end{abstract}


\section{Introduction}\label{introduction}

Dwarf galaxies are, by number, the most numerous systems in the Universe. Under the currently accepted $\Lambda$ Cold Dark Matter ($\Lambda$CDM) paradigm, these dwarfs constitute the buildings blocks of larger galaxies (see for example \citealt{1978ApJ...225..357S}; \citealt{1978MNRAS.183..341W}; \citealt{2006MNRAS.371..885R}).
Although $\Lambda$CDM explains many phenomena very well, it overpredicts the number of bound dark matter haloes in the Local Volume of galaxies as compared to the number of visible galaxies (e.g. \citealt{1999ApJ...522...82K}; \citealt{1999ApJ...524L..19M}).
Several mechanisms have been proposed to solve this so-called \textquote{missing satellites} problem (e.g. \citealt{2010MNRAS.402.1995M}; \citealt{2016MNRAS.457.1931S}; \citealt{2019MNRAS.487.5799R}). These include internal feedback processes like stellar winds and supernovae that suppress star formation (\citealt{1986ApJ...303...39D}), and external processes like reionisation (e.g. \citealt{1992MNRAS.256P..43E}), galactic tides (e.g. \citealt{2006MNRAS.366..429R, 2006MNRAS.367..387R, 2007MNRAS.378..353K}) and ram-pressure stripping (e.g. \citealt{2006MNRAS.369.1021M}; \citealt{2009ApJ...696..385G}; \citealt{2013MNRAS.433.2749G}) that quench star formation. The nearby Magellanic Clouds (MCs) -- an interacting pair of dwarfs in the process of falling into the Milky Way (MW, \citealt{2013ApJ...764..161K}) -- provide a natural Rosetta Stone for probing all of the above processes in intricate detail \citep[e.g.][]{2017MNRAS.471.4571C}.

The Small and Large Magellanic Cloud (SMC/LMC) lie at a distance of  $\sim$62\,kpc and $\sim$50\,kpc, respectively \citep{2016ApJ...816...49S,2019Natur.567..200P}. A plethora of evidence points to a recent interaction between the LMC and the SMC that resulted in the formation of the purely gaseous Magellanic Stream (\citealt{2008ApJ...679..432N}; \citealt{2016ARA&A..54..363D}), the Magellanic Bridge  (\citealt{1974ApJ...190..291M}; \citealt{2013ApJ...768..109N}; \citealt{2015MNRAS.452.4222N};  \citealt{2017MNRAS.471.4571C}) -- that comprises a stellar and gas arch connecting both galaxies --, and the Leading Arm (first identified by \citealt{1998Natur.394..752P}), a stream of gas on the opposite side of the Magellanic Stream. Another interesting feature showing the past interactions of the Clouds is the RR Lyrae overdensity found by \citet{2017MNRAS.466.4711B}, although these results are put into question by \citet{2020ApJ...889...26J} who were not able to obtain an evident connection between the MCs without many spurious sources in the sample.

In recent decades, the accurate measurements of the MCs' proper motions have considerably increased our understanding of the Magellanic system (\citealt{2009AJ....137.4339C}; \citealt{2013ApJ...764..161K}). Their proper motions indicate that the MCs are most likely on their first infall into the MW, or on a long period orbit, depending on the assumed mass of the MW and the LMC (\citealt{2007ApJ...668..949B}; \citealt{2017MNRAS.464.3825P}). This has become the new paradigm driving our understanding of the MCs' evolution.
 
From a theoretical point of view, the new proper motion data, together with improved estimates of their dynamical masses, gas fractions and stellar disc scale-lengths, have allowed for the advent of increasingly realistic dynamical simulations of the MCs (e.g. \citealt{2010ApJ...721L..97B, 2012MNRAS.421.2109B}; \citealt{2012ApJ...750...36D}). $N$-body simulations reproduce well the Magellanic Stream and Leading Arm, suggesting that they result from a tidal interaction between the Clouds \citep{2006MNRAS.371..108C}. Recently, \citet{2019MNRAS.488..918T} argue that the formation of the Leading Arm is present in models without a
corona but it is inhibited by the hydrodynamic interaction with the hot component raising further questions on the nature of this feature.  Chemodynamical simulations (\citealt{2009PASA...26...48B}; \citealt{2012ApJ...750...36D}) show that the Magellanic Bridge and the \textquote{Counter-Bridge} (a minor structure extending away from the SMC) also formed as a result of LMC-SMC interactions. Finally, using hydrodynamical simulations, \citet{2018MNRAS.480L.141A} propose a new scenario to explain the counter-rotating stellar population in the LMC in which this is the result of a minor retrograde merger with another dwarf galaxy more than 3\,Gyr ago.

From an observational perspective, the past decade has witnessed the discovery of many new structures around the MCs, including a slew of `ultra-faint' dwarfs likely to be associated to the MCs (\citealt{2015ApJ...813..109D}; \citealt{2015ApJ...805..130K}; \citealt{2015MNRAS.453.3568D}; \citealt{2016ApJ...833L...5D}; \citealt{2016MNRAS.461.2212J}; \citealt{2017MNRAS.465.1879S}; \citealt{2018ApJ...852...68C}; \citealt{2018ApJ...867...19K}; \citealt{2019arXiv190709484E}); a `stellar cloud' within the LMC tidal radius \citep{2016MNRAS.459..239M}; SMC structures such as the SMCNOD that could owe to a primordial SMC satellite in advanced stage of disruption \citep{2017MNRAS.468.1349P} and the shell-like overdensity studied by \cite{2019A&A...631A..98M} that could owe to a recent star formation event triggered by interactions between the MCs and/or the MW; and the very extended, low-density, envelope of stellar material with a disturbed shape around the LMC \citep{2019ApJ...874..118N}. All of these structures can be used as tracers of the gravitational interaction history of the MCs \citep{2016ApJ...825...20B}.

Historically, the LMC has been studied in greater detail than its smaller companion. This owes to several factors, such as the slightly lower distance from us, the more defined structure as a barred irregular \citep{2017AAS...22941605C}, its less perturbed nature, and its larger mass (\citealt{2002AJ....124.2639V}; \citealt{2016MNRAS.456L..54P}; \citealt{2018ApJ...866...90C}; \citealt{2019MNRAS.487.2685E}). Proper motion studies show that the LMC's disc has a clockwise rotation in the plane of the sky, suggesting an inner region that is relatively \textquote{unperturbed} (\citealt{2014ApJ...781..121V}; \citealt{2016A&A...595A...1G}). The periphery of the LMC, however, shows evidence of more complicated substructures indicative of tidal stripping by the MW (\citealt{2018ApJ...858L..21M}; \citealt{2019ApJ...874..118N}; \citealt{2019MNRAS.482L...9B}) and of interactions with the SMC, as shown by \citep{2016MNRAS.456..602B} who studied the distribution of Blue Horizontal Branch (BHB) stars.

The SMC has a more intricate structure than the LMC with complex dynamics and kinematics that, in spite of its vicinity to us, are not yet fully understood. It is more metal poor than the LMC \citep{2008AJ....136.1039C}, has an old stellar population that is older than its single old globular cluster (\citealt{2007AJ....133.2037N}; \citealt{2008AJ....136.1703G}), and presents a peculiar star formation history with conspicuous enhancements at various epochs \citep{2009ApJ...705.1260N}. 

 Studying the kinematics of LMC supergiants, \citet{2011ApJ...737...29O} found metallicity differences in more than 5\% of the population, providing strong evidence that the kinematically distinct population originated in the SMC. Measuring radial velocities of intermediate-age red giant  branch (RGB) stars, \citet{2017MNRAS.471.4571C} found that these stars were actually an SMC population, unequivocally confirming  the tidal  stripping  scenario in  the Magellanic  Bridge  region.
 All these findings constitute a clear evidence of the SMC being tidally stripped by the LMC.

The SMC-LMC interactions not only affect the kinematics but also the star formation rates (SFR)
as shown by photometric studies of the stellar populations in the outskirts of the SMC where similar evidences for tidal effects are found in the form of enhanced SFR $\sim$200 Myr ago (\citealt{2013ApJ...768..109N}; \citealt{2014MNRAS.442.1897R}; \citealt{2015MNRAS.452.4222N}). 
Massana et al. (2020, in prep.) found a break in the young stellar mass profile, indicating that the SMC has a disturbed star-forming outer ring while the old underlying population presents a smoother profile.
\citet{2016A&A...591A..11D}, studying the ages of star clusters throughout the SMC, found clear age and metallicity gradients consistent with tidal interactions between the LMC and SMC. These results are supported by numerical models of the Magellanic system that predict that the SMC is likely to show coherent rotation only at large radii \citep{2012MNRAS.421.2109B}.
A comprehensive study of the classical Cepheids in the SMC \citep{2017MNRAS.472..808R} showed a complex geometric structure, with the near side forming a rough spheroidal shape before gradually shifting into a more linear shape, adding more detail to the SMC but presenting yet another potentially conflicting stellar structure. 

In spite of the above evidences for tidal effects, the extent of the SMC disruption is still uncertain. Evidence of the tidal interaction between the Clouds affecting inner regions of the SMC and not just the outskirts is still lacking. Thus, the internal kinematics and structure of the SMC remain puzzles to be solved.
\citet{2004ApJ...604..176S}, and more recently \citet{2019MNRAS.483..392D}, claim to have recovered a strong signal of ordered rotation in the $\mbox{H\,{\sc i}}$ gas despite the disruptions the SMC has been through while \citet{2019ApJ...887..267M} dispute this claim. \citet{2014MNRAS.442.1663D} was able to recover a rotation signal in the stellar component (albeit with significant uncertainties in the parameters of their disc model), while \citet{2018A&A...616A...1G} found no rotation signal, and \citet{2018ApJ...864...55Z} found some evidence of tidal disruption in the innermost regions. \citet{2014MNRAS.442.1663D} also found signs of tidal stripping in the outer regions of the SMC ($\sim$4 kpc) and a velocity gradient along the northwest-southeast axis. The analysis of RGBs in the inner region ($\sim$2 kpc) conducted by \citet{2006AJ....131.2514H} showed that this older population was dynamically separated from the $\mbox{H\,{\sc i}}$ gas, with a very weak rotation signature and a significant velocity dispersion ($\sigma=27.5\pm0.5$\,km\,s$^{-1}$) suggesting that the SMC possess a spheroidal structure, rather than a disc-like one. Studying young massive stars, \citet{2008MNRAS.386..826E} found a velocity dispersion of 28\,km\,s$^{-1}$, very similar to the one reported by \citet{2014MNRAS.442.1663D} who analysed a sample of RGB stars. While it is not surprising to find a dispersion-supported old stellar population in a rotationally-supported gas disc, it is puzzling for the young stars to be similarly decoupled from the gas kinematics from which they were recently born.

Distance estimations using Classical Cepheids (e.g. \citealt{2016AcA....66..149J} and \citealt{2017MNRAS.472..808R}) report evidences for an ellipsoidal morphology of the SMC that appears to be very extended along the line of sight, with different depths for different regions and a mean depth of $\sim$5\,kpc (rising to and above $\sim$20\,kpc in some regions). Using RR Lyrae as distance tracers there are observations of similar morphology and depth (e.g.  \citealt{2017AcA....67....1J} and \citealt{2018MNRAS.473.3131M}) with the SMC's bound structure identified as an ellipsoid with an axes ratio of 1:1.1:3.3 with the main axis roughly along the line of sight. These observations are at odds with the increasing amount of evidence for tidal disruption that would disfavor such an extended bound structure of the SMC.

With the goal of shedding light on the internal structure and dynamics of the SMC, we present here a kinematic study using $\sim$6000 RGB stars in the SMC (from our own observations and those of \citealt{2014MNRAS.442.1663D}) up to $\sim5\degr$ from the photometric centre, obtained using the AAOmega spectrograph \citep{2004SPIE.5492..389S}. We combine this with data from the \textit{Gaia} Data Release 2 \citep[DR2][]{2018A&A...616A...1G}. In Section ~\ref{observ} we present the observational data. In Section ~\ref{RadVels} we discuss the radial velocity estimations. Section ~\ref{propmot} covers the cross-match with the \textit{Gaia}~DR2 catalogue and the proper motion analysis. In Section ~\ref{discussion} we present the discussion of our results. Finally, our conclusions are presented in Section ~\ref{conclusions}.

\section{Observational Material}\label{observ}

\subsection{Target selection}\label{targets}

\begin{figure*}
\includegraphics[width=\textwidth]{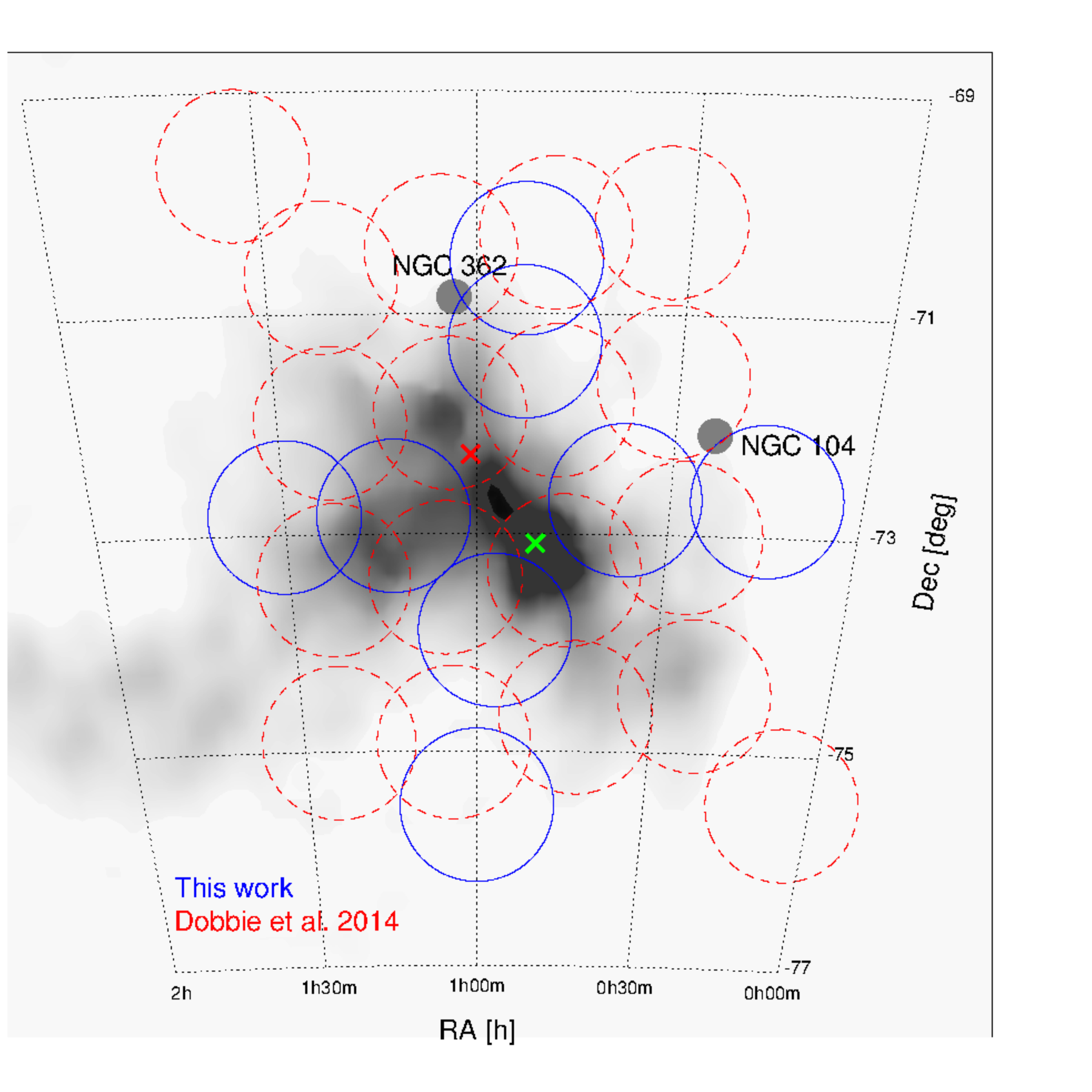}
\caption{The SMC region in a Hammer equal area projection. The greyscale contours represent the $\mbox{H\,{\sc i}}$ emission integrated over the velocity range 80 $\leq$ V [km\,s$^{-1}$] $\leq$ 200, where each contour represents the $\mbox{H\,{\sc i}}$ column density taken from the Leiden/Argentine/Bonn (LAB) survey of Galactic $\mbox{H\,{\sc i}}$ \citep{2015A&A...578A..78K}. Blue solid circles are the regions observed in this work, red dashed circles are those regions observed by \citet{2014MNRAS.442.1663D}. As reference we report the SMC photometric centre derived by the sample from \citet{2002AJ....123..855Z} and the $\mbox{H\,{\sc i}}$ gas kinematic centre by \citet{2019MNRAS.483..392D}, respectively as green and red crosses.\label{fig:spatial}}
\end{figure*}

\begin{figure}

\includegraphics[width=\columnwidth]{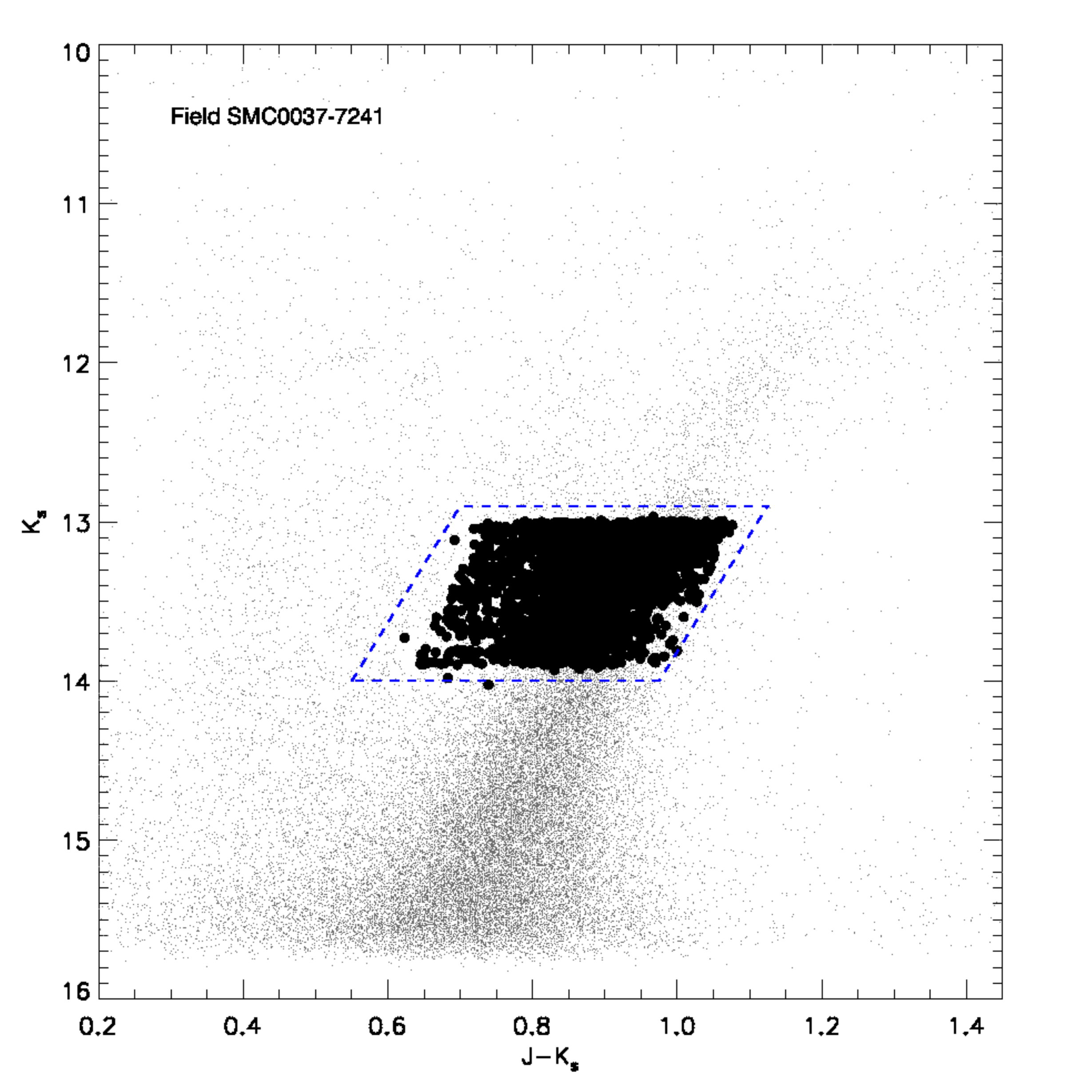}
\caption{Location of the region used to select the target stars over-plotted to the $J - K_S$ versus $K_S$ colour-magnitude diagram for one of the fields. The black points inside the blue dashed parallelogram are the stars sampled with AAOmega.\label{fig:dcm_sel}}
\end{figure}

With the goal of sampling different position angles (PAs) and radii from the SMC centre, we observed the eight SMC fields shown as solid blue circles in Fig.~\ref{fig:spatial}. The greyscale contours in the same image represent the HI column density taken from \citet[][see the caption for further details]{2015A&A...578A..78K}. The centres of the selected fields are listed in Table~\ref{fields_coordinates}. Our data is complementary to those from 
 \citet[][red dashed circles in Fig.~\ref{fig:spatial}]{2014MNRAS.442.1663D}, who performed a similar study but covered mainly the SMC body. We sampled regions between the fields from \citet{2014MNRAS.442.1663D} and went slightly further in some directions. The spectroscopic targets were selected from the expected position of the upper RGB in the colour-magnitude diagram (CMD) from the Two Micron All Sky Survey \citep[2MASS;][]{2006AJ....131.1163S}. This is shown in Fig.~\ref{fig:dcm_sel} where we present the  ($J - K_S$, $K_S$) CMD for one of the fields (see Table~\ref{fields_coordinates}) with the targets overlapped.  

\begin{table}
 \centering
\caption{Coordinates of the centres of each field observed\label{fields_coordinates}} 
 \begin{tabular}{@{}lcc@{}}
\hline
Field Name  & RA & Dec \\
   \hline
SMC0015\_7240 & 00:15:00.00 & -72:40:00.0 \\
SMC0037\_7241 & 00:37:00.00 & -72:41:00.0 \\
SMC0053\_7030 & 00:53:00.00 & -70:30:00.0 \\
SMC0053\_7115 & 00:53:00.00 & -71:15:00.0 \\
SMC0057\_7353 & 00:57:00.00 & -73:53:00.0 \\
SMC0100\_7530 & 01:00:00.00 & -75:30:00.0 \\
SMC0113\_7250 & 01:13:00.00 & -72:50:00.0 \\
SMC0130\_7250 & 01:30:00.00 & -72:50:00.0 \\
\hline
\end{tabular}
\end{table}

\subsection{Optical spectroscopy}\label{spectro}

The spectroscopic observations were secured on the nights of the 2nd and 3rd October 2011 and 10th, 11th and 12th November 2014 with the AAOmega spectrograph \citep{2004SPIE.5492..389S} fed by the Two Degree Field (2dF) multi-object system installed at the prime focus of the Anglo-Australian Telescope located at Siding Spring Observatory (Australia)\footnote{Program I.D.: ATAC/2012B/010, and S/2014B/04}. The 2dF+AAOmega is a dual-beam spectrograph that allows to allocate up to 400 2"-size optical fibres within a 2\degr\ field of view. In the red arm, we used the grating 1700D centred on $\sim$8500\,\AA, providing a spectral resolution of $R\sim 8500$. For each field we acquired three exposures of 3600~s. In total, we obtained spectra with a signal-to-noise ratio (SNR) per pixel larger than 5 for 2573 stars.

The initial steps of the data reduction, including bias subtraction, flat-field normalisation, fiber tracing and extraction, and wavelength calibration, were performed with the dedicated 2dF data reduction pipeline \citep[2dfdr\footnote{See https://www.aao.gov.au/science/software/2dfdr}; ][]{2010PASA...27...91S}. Our own software was used to subtract the emission sky lines following the procedure described in detail by \citet{2017MNRAS.470.4285C}. Briefly, the scale factor that minimises the sky line residuals is determined by comparing the spectrum observed in each fiber with a master sky spectrum obtained by averaging the spectra of the nearest ten fibers placed on sky positions without any source.

\section{Radial Velocities}\label{RadVels}

\subsection{Radial velocity determination}\label{RadVelsDet}

After applying the heliocentric correction, individual spectra were averaged to obtain the combined spectrum for each star using the individual SNR as weight and an average sigma clipping rejection algorithm to remove deviant pixels. Those individual exposures with very low SNR, $<$5 pix$^{-1}$, are rejected. The resulting \textquote{combined} spectrum is interpolated with each individual exposure to find and minimise small shifts between them. This procedure is repeated until the convergence criterion of shifts lower than $10^{-3}$ per pixel is met. The combined spectrum is then cross correlated with a grid of synthetic spectra. The details about the computation of this grid can be found in \citet{2018A&A...618A..25A}. This grid has three dimensions: metallicity, [M/H]; effective temperature, $T_{\rm eff}$; and surface gravity, log $g$.
The metallicity ranges from -5.0 to +1.0\,dex with a step of 0.5\,dex, the temperature is in the range 3\,500 $\leq$ $T_{\rm eff}[K]$ $\leq$ 6\,000 with a step of 500\,K. Finally, the range in gravity covers log$g=$ 0.0 to 5.0\,dex with a step of 1.0\,dex. This implies that the grid contains 432 synthetic spectra. For the $\alpha$-elements abundances the spectra were computed assuming [$\alpha$/Fe]=0.5\,dex for [Fe/H]$\leq-1.5$\,dex, [$\alpha$/Fe]=0.0\,dex for [Fe/H]$\geq$+0.0\,dex, and linear between them. The abundances of other elements were fixed to the Solar values \citep{2005ASPC..336...25A} and the microturbulence velocity was fixed to 1.5\,km\,s$^{-1}$. The shifts previously obtained between the combined spectrum and each individual exposure are applied to the radial velocity determined from the cross correlation with the grid to obtain individual values for each exposure. Finally, these individual values are averaged to derive the final radial velocity using the single exposures' SNRs as weights.

\begin{table}
 \centering
\caption{Range of each dimension in the synthetic grid used for the radial velocity determination.\label{griddimension}} 
 \begin{tabular}{@{}lccc@{}}
\hline
Dimension  & First & Last & Step \\
   \hline
$[$M/H$]$ (dex)& -5.0 & +1.0  & 0.5\\
$T_{\rm eff}$ (K) & 3500  & 6000  & 500\\
log $g$ (dex) & 0.0  &  5  & 1.0 \\
\hline
\end{tabular}
\end{table}

\begin{figure}
\includegraphics[width=\columnwidth]{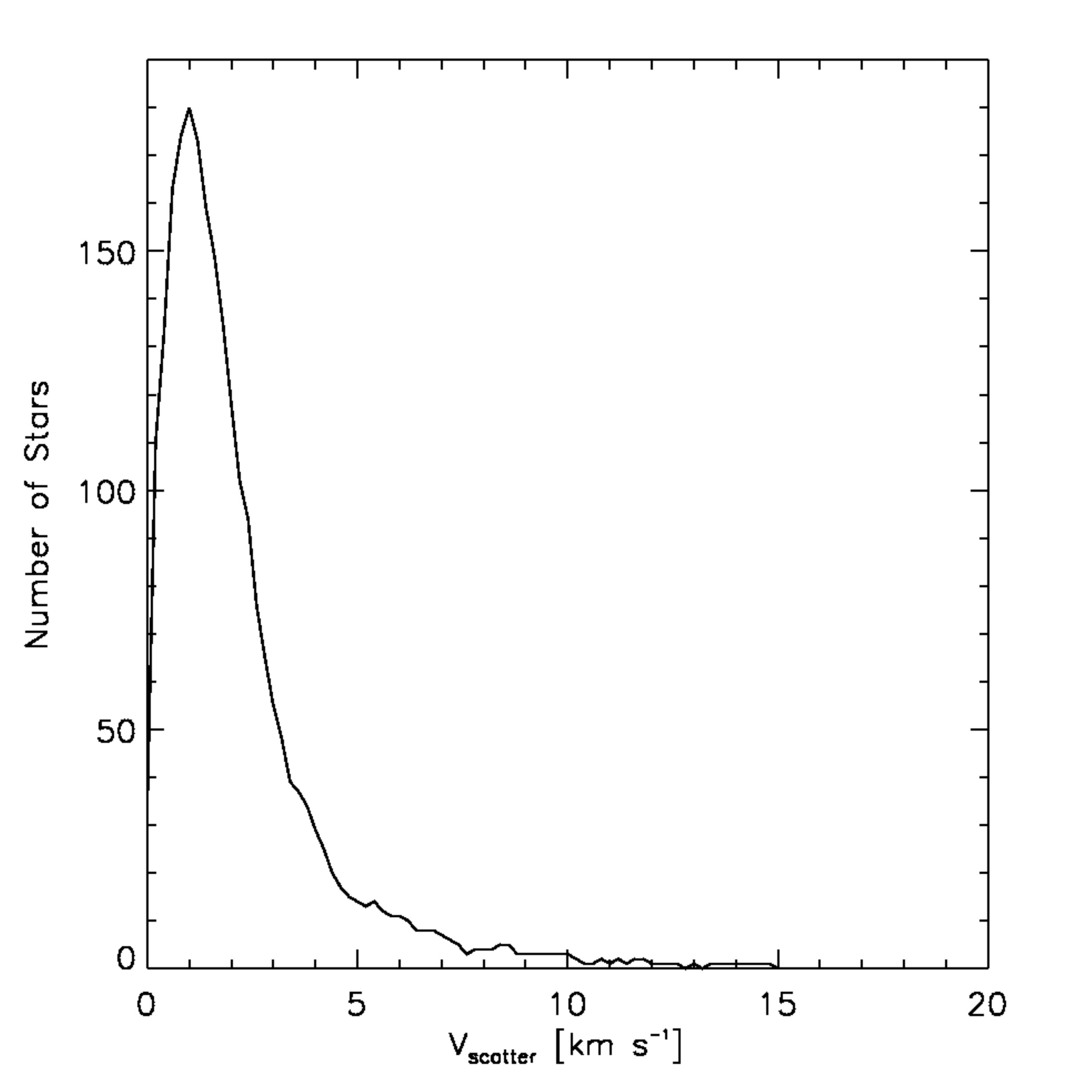}
\caption{Distribution of the radial velocity scatter for those stars with at least 3 individual exposures.\label{fig:vscatter}}
\end{figure}

\subsection{Radial velocity uncertainties}\label{RadVelunc}

The procedure followed to derive the radial velocities allows us to also determine the scatter of the multiple individual radial velocity determination, $V_{\rm scatter}$. This provides a better estimation of the internal precision than the typical uncertainty determined from the cross-correlation peak. Fig.~\ref{fig:vscatter} shows the distribution of the radial velocity scatter obtained for those stars with at least three individual measurements. The distribution has a mode of 1.0 km\,s$^{-1}$ and about 90\% of the objects have $V_{\rm scatter}\leq$ 5\,km\,s$^{-1}$ and only 2.5\% of the stars have values larger than 10\,km\,s$^{-1}$. Large $V_{\rm scatter}$ values may suggest that the star is in a binary system. For this reason, we have removed from our analysis all stars with $V_{\rm scatter}>$10\,km\,s$^{-1}$.

\begin{figure}
\includegraphics[width=\columnwidth]{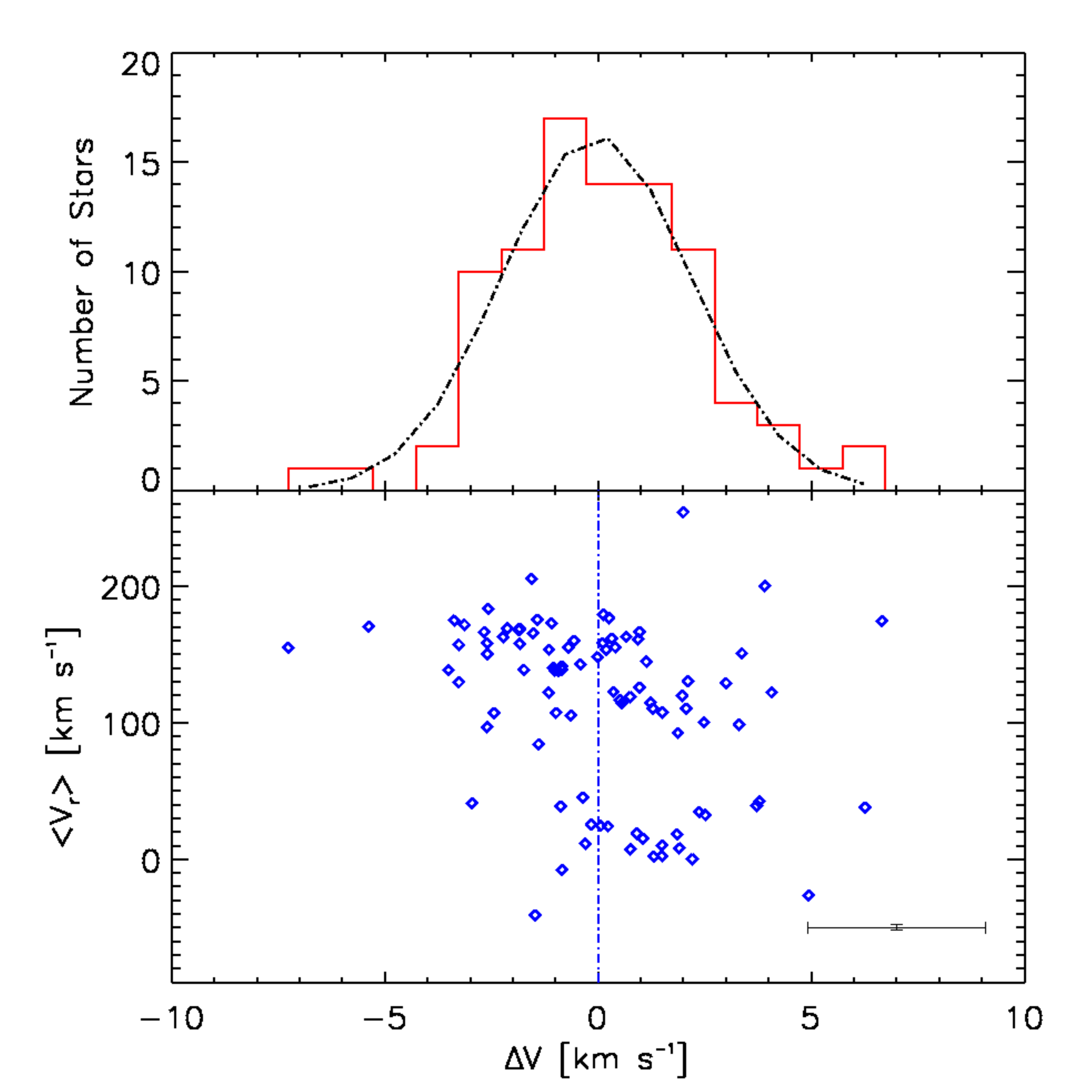}
\caption{{\it Top panel}: distribution of radial velocity differences for the sample of 91 internal cross matches binned at 1\,km\,s$^{-1}$ intervals (red histograms) and fitted with a Gaussian (black dash-dotted line). {\it Bottom panel}: dispersion of the radial velocity differences of the same sample against the computed mean radial velocity for each star (blue dash-dotted line is the zero difference line). The typical error bar is reported in the bottom right corner. \label{fig:comp_int}}
\end{figure}

The 91 stars that have been observed in common in two different fields are useful to estimate the internal accuracy of our radial velocity determination. We determined the radial velocity for these stars in each field independently. This is depicted in Fig.~\ref{fig:comp_int} where the bottom panel shows the difference of the radial velocities obtained in each case as a function of average velocity from all the measurements and the top panel shows the distribution of the differences. As seen from Fig.~\ref{fig:comp_int}, there is no clear trend with the radial velocity (bottom panel) and the difference in the radial velocity distribution presents the expected Gaussian behavior (top panel). The Gaussian fit to the error distribution is centred at 0.4$\pm$0.2 km\,s$^{-1}$ with a $\sigma$ of 1.1$\pm$0.2 km\,s$^{-1}$ and a median of 0. This means that our analysis is robust and that the observed dispersion is within the expected uncertainty taking into account the spectral resolution used.

\subsection{Comparison with the literature}\label{RadVelcomp}

The SMC has been the target of several studies that have derived radial velocities for different objects. In particular, some of them have targeted late type RGBs as in our case. The comparison of our sample with those studies not only allows us to put our sample in the global context of the SMC kinematic analysis but also provides an external comparison of our radial velocity measurements. To achieve this we cross-matched our sample with the ones available in the literature covering the same portion of the sky, finding objects in common with the Tool for OPerations on Catalogues And Tables \citep[\textit{TOPCAT}; ][]{2005ASPC..347...29T}. Table~\ref{rvliterature} summarises the studies we compared with, the stars in common, and the differences between the radial velocities determined in each case.

\begin{table}
 \centering
\caption{Compilation of various radial velocity determinations from the literature.\label{rvliterature}} 
 \begin{tabular}{@{}lcccc@{}}
\hline
Study & Cross & $\langle\Delta V\rangle$ & $\sigma_{\Delta V}$ & $\widetilde{\Delta V}$\\
 & matches & km\,s$^{-1}$ & km\,s$^{-1}$& km\,s$^{-1}$\\
   \hline
\citet{2006AJ....131.2514H} & 11 & $1\pm2$ & $8\pm2$ & 0 \\
\citet{2008AJ....136.1039C} & 8 & $-17\pm5$ & $13\pm4$ & -15 \\
\citet{2010AJ....139.1168P} & 0 & - & - & - \\
\citet{2010ApJ...714L.249D} & 92 & $-19\pm2$ & $20\pm2$ & -16 \\
\citet{2014MNRAS.442.1663D} & 175 & $1.2\pm0.1$ & $1.6\pm0.1$ & 1 \\
\hline
\end{tabular}
\end{table}

\citet{2006AJ....131.2514H} sampled 2046 red giant stars in the central  4\,kpc $\times$ 2\,kpc of the SMC. Despite the fact that our study includes some more external parts of the galaxy there are a few stars in common between both works and our radial velocity determinations are in agreement within the uncertainties.

\citet{2008AJ....136.1039C} obtained infrared $\mbox{Ca\,{\sc ii}}$ triplet spectroscopy in over 350 stars of the RGB spread in several fields of the SMC body with the main goal of deriving stellar metallicities. The few stars in common between our sample and their sample show a significant difference between the radial velocities determined by each work. The \citet{2008AJ....136.1039C} data were acquired with FORS2 on the VLT with a low resolution ($<$5000). FORS2 is a multi-slit spectrograph where misalignments between stars and slits hampered the determination of the radial velocities. This offset can produce significant differences in radial velocities that may explain the disagreement with our determinations \citep[as reported for example by][]{2002MNRAS.336..643I}.

\citet{2010ApJ...714L.249D} derived radial velocities for 683 red giant stars in 10 SMC fields using the Hydra multi-fiber spectrograph (with a resolution of $\sim$4200) mounted on the V.M. Blanco 4m telescopo at Cerro Tololo (Chile). Their results for 92 stars in common with our sample show a marked disagreement with our results. There is no clear explanation for this discrepancy that might stem from the lower resolution of the Hydra spectrograph and other differences between the two instrumental setups.

\begin{figure}
\includegraphics[width=\columnwidth]{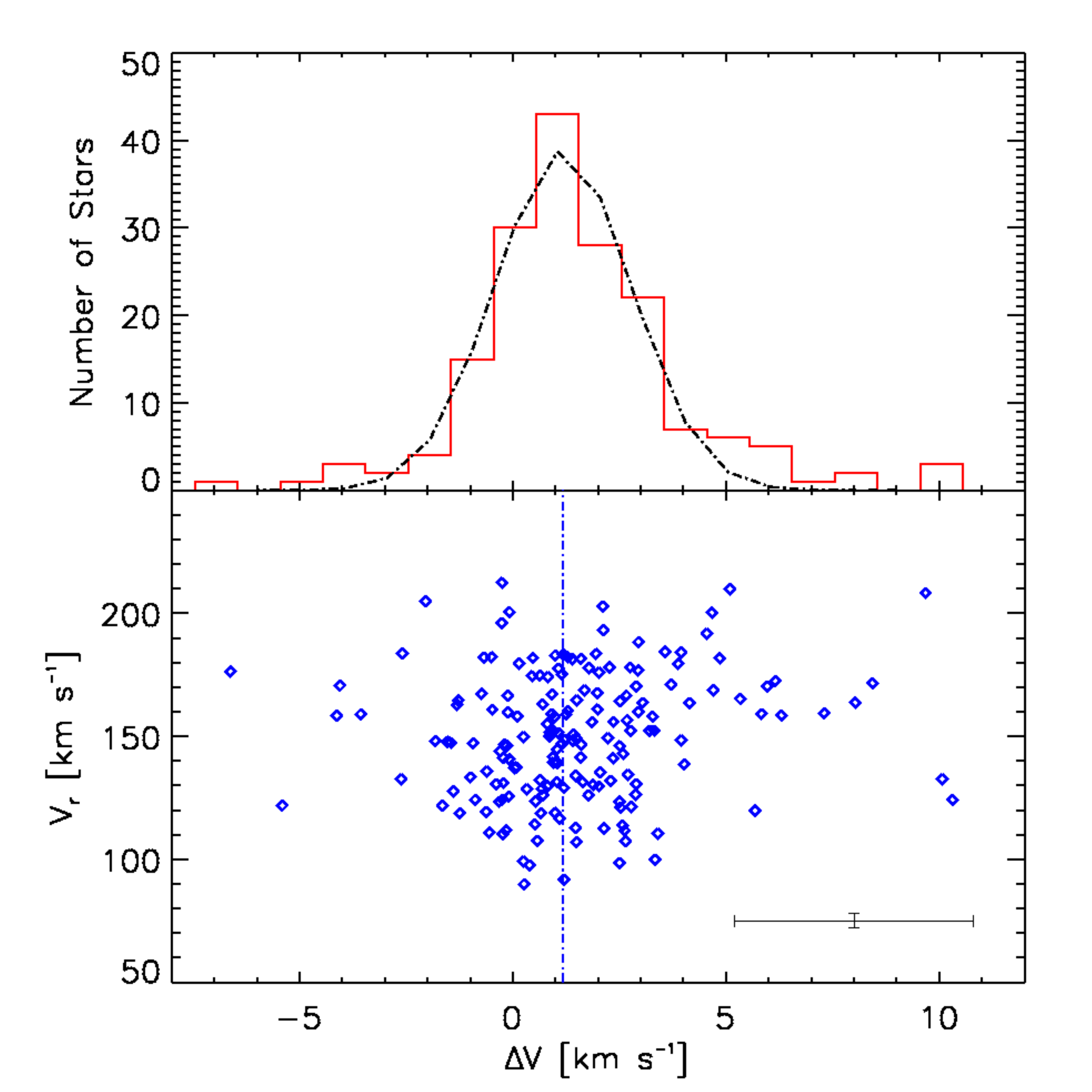}
\caption{{\it Top panel}: distribution of radial velocity differences for the sample of 175 stars in common between our data set and that of \citet{2014MNRAS.442.1663D} binned at 1\,km\,s$^{-1}$ intervals (red histograms) and fitted with a Gaussian (black dash-dotted line). {\it Bottom panel}: dispersion of the radial velocity differences of the same sample against our computed radial velocity for each star (blue dash-dotted line is the mean value of the radial velocity differences). The typical error bar is reported in the bottom right corner. \label{fig:comp_Dobbie}}
\end{figure}

Finally, \citet{2014MNRAS.442.1663D} derived radial velocities for more than 4000 red giant stars in 18 SMC fields mostly located in the centre of the galaxy and along the major and minor axes. They also used the 2dF+AAOmega spectrograph at the Anglo-Australian Telescope with our same instrumental configuration as well as a relatively similar approach to obtain the radial velocities. We found 175 stars in common with this data set, giving us a sample big enough to run some more proper statistics on the cross matches. Fig.~\ref{fig:comp_Dobbie} shows the distribution (top panel) and dispersion (bottom panel) of the radial velocity differences between the two analysis. The Gaussian fit to the distribution is centred at 1.2$\pm$0.1\,km\,s$^{-1}$ with a dispersion of 1.6$\pm$0.1\,km\,s$^{-1}$ and with a median of 1\,km\,s$^{-1}$. This result proves the agreement between our analysis and that of \citet{2014MNRAS.442.1663D} and the consistency between the two data sets. Owing to the facts that both samples have been derived with the same instrumental configuration and we have proved that the radial velocities derived in each case are compatible, we created an \textquote{extended} sample resulting from the combination of our data and that of the confirmed SMC RGB stars from \citet{2014MNRAS.442.1663D}. This extended sample not only provides a better spatial coverage of the SMC body out to $\sim4$\,kpc from the photometric centre but also raises the statistical robustness of all the derived SMC parameters due to the higher number of stars.  

There are other works in the literature that have sampled a significant number of objects in the SMC such as \citet{2010AJ....139.1168P} but we cannot compare our analysis with theirs given that there are no stars in common.

\subsection{SMC's bulk radial velocity along the line of sight}\label{bulkVel}

\begin{figure}
\includegraphics[width=\columnwidth]{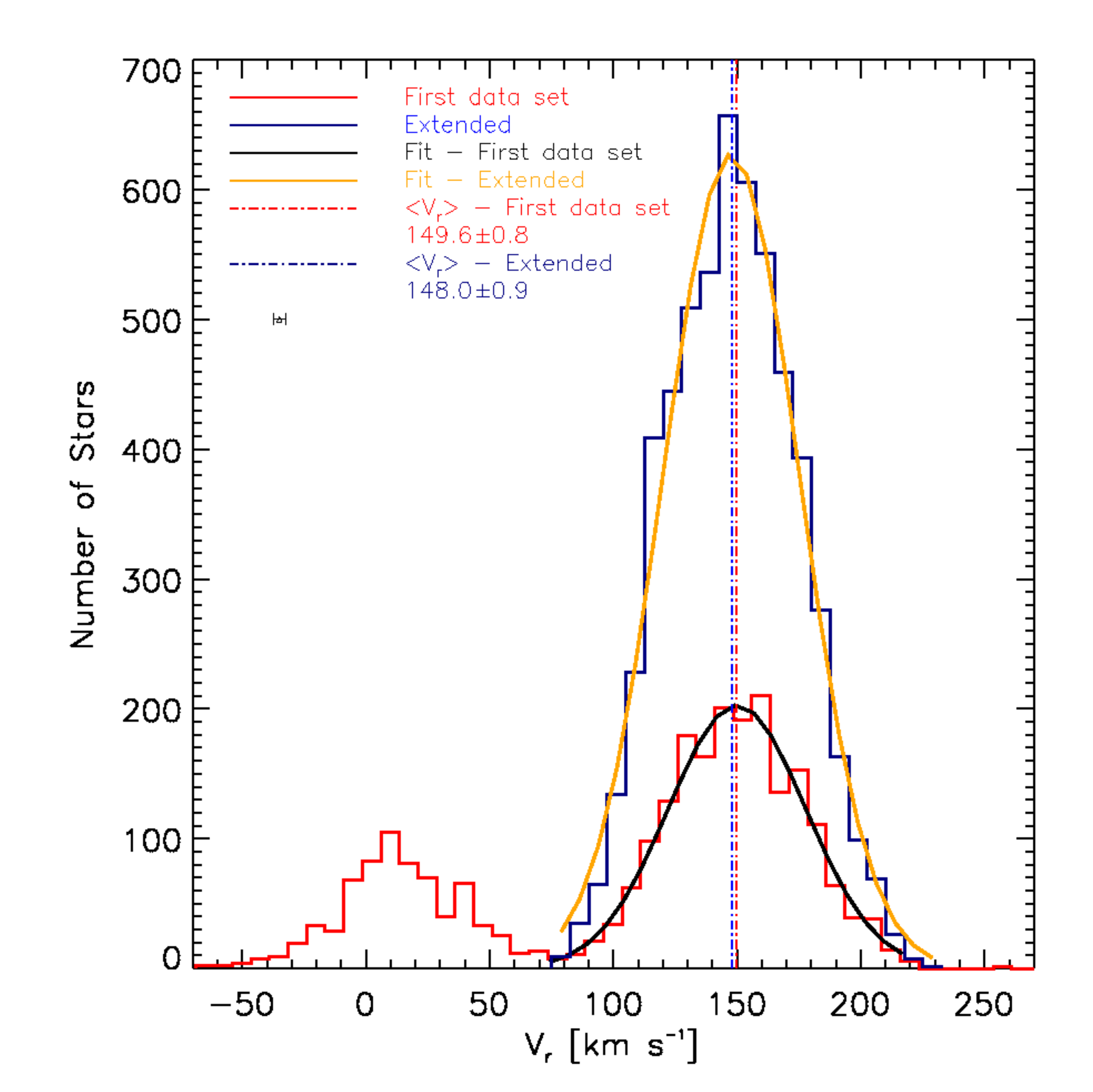}
\caption{Distribution of radial velocities in our data set (red histogram) and in the extended data set (blue histogram), binned at 7.5\,km\,s$^{-1}$ intervals, and fitted Gaussians (respectively black and orange lines). The blue and red histograms are perfectly overlapping in the area of the secondary peak around 10\,km\,s$^{-1}$ because the data taken from \citet{2014MNRAS.442.1663D} include only confirmed SMC members (i.e. with 70 < $V_r$ < 230 km\,s$^{-1}$). The dash-dotted lines indicate the mean value of each corresponding distribution (shown in the same colours), with the values reported in the legend. A point showing the mean error in the individual radial velocities is shown below the legend. \label{fig:rv_dist}}
\end{figure}

Fig.~\ref{fig:rv_dist} shows the radial velocity distribution of the observed stars in our data set (red histogram) and in the extended data set (blue histogram). The distributions show two clear peaks: one around 10 km\,s$^{-1}$ and the other one around 150 km\,s$^{-1}$. The first is due to foreground Milky Way stars in the line of sight of the SMC, while the latter corresponds to the SMC stars as expected from previous studies. The first peak is identical for both samples since we took only the confirmed SMC members from \citet{2014MNRAS.442.1663D}, hence the blue histogram is not seen in this part of the plot. In our data set, we have 1861 stars with radial velocity between 70 and 230 km\,s$^{-1}$, the extended data set has $\sim$5700 stars with radial velocity in the same range. This is the expected range of radial velocities for stars belonging to the SMC and in the following analyses we will focus on these stars and ignore the field stars.

To characterise the radial velocity distributions in Fig.~\ref{fig:rv_dist} we used a Markov Chain Monte Carlo \citep[MCMC; ][]{2013PASP..125..306F} maximum likelihood estimator commonly used \citep[for example by][Eq.~\ref{likelihood}]{2017ApJ...838....8L} to fit the confirmed SMC stars radial velocity distribution. In Eq.~\ref{likelihood} $< V_r >$ and $\sigma_{vr}$ are respectively the mean radial velocity and the dispersion of the data set and $v_i$ and $\sigma_i$ are respectively the radial velocity and its error for each individual star:
\begin{equation}\label{likelihood}
	\ln{\mathcal{L}}=-\frac{1}{2}\Bigg[\sum_{i=1}^{N} \ln{(\sigma_{vr}^2+\sigma_i^2)}+\sum_{i=1}^{N} \frac{(v_i-< V_r >)^2}{\sigma_{vr}^2+\sigma_i^2}\Bigg]
\end{equation}
The $< V_r >$ and $\sigma_{vr}$ recovered through the MCMC maximum likelihood estimation are the ones used to generate the Gaussian fits to the radial velocity distribution of the SMC stars plotted in Fig.~\ref{fig:rv_dist} and are reported in Table~\ref{rv_smc_lit} with comparisons from the literature. As expected, our results are in good agreement with those from \citet{2014MNRAS.442.1663D} given that the data was taken using the same instrument and similar spatial distribution. The consistency of results, as already proved in the case of individual radial velocity determinations, allows us to merge the two data sets into the extended one. The small discrepancy with the results from \citet{2006AJ....131.2514H} can be explained by the fact that we sampled different regions of the SMC and in fact the discrepancy lessens when we analyse the data of the extended data set that has a more complete coverage of the SMC.
Our results are roughly consistent with those that can be derived by \citet{2015AJ....149..154P} in which they observed a small sample of $\sim$100 stars from various clusters in the main body and the outskirts of the SMC with FORS2 on the VLT. As with \citet{2008AJ....136.1039C}, the discrepancy in results may stem from the low resolution and misalignment problems of FORS2, but in this case the authors were aware of the limitations of the instrument and modified their analysis pipeline to correct in part for these effects, thus our results are compatible with theirs.

\begin{table}
\centering
\caption{SMC Radial Velocity and Radial Velocity Dispersion, the listed uncertainties include both statistical and systematic errors \label{rv_smc_lit}} 
 \begin{tabular}{@{}lcc@{}}
\hline
Study & RV & $\sigma$ \\
 & km\,s$^{-1}$ & km\,s$^{-1}$ \\
   \hline
Our data set & 149.6$\pm$0.8 & 26.9$\pm$0.5 \\
\citet{2014MNRAS.442.1663D} & 147.8$\pm$0.5 & 26.4$\pm$0.4 \\
``Extended'' & 148.0$\pm$0.9 & 25.6$\pm$0.2 \\
\citet{2006AJ....131.2514H} & 145.6$\pm$0.6 & 27.6$\pm$0.5 \\
\citet{2015AJ....149..154P} & 149.3$\pm$3 & 24.2$\pm$3.2 \\
\hline
\end{tabular}
\end{table}

\subsection{Radial velocity map: gradients and features}\label{RadVel_features}

In order to check for signs of a gradient or features in our extended sample, we re-analyse our set of radial velocities allowing for a velocity gradient as in \citet{2017MNRAS.467..573C} using the likelihood function of Eq.~\ref{likelihood-bis} with the systemic velocity $< V_r >$, the velocity dispersion $\sigma_{vr}$, the velocity gradient $\frac{\mathrm{d}v}{\mathrm{d}\chi}$ and the PA of the gradient $\theta$ as the parameters to be fit:
\begin{equation}\label{likelihood-bis}
	\ln{\mathcal{L}}=-\frac{1}{2}\Bigg[\sum_{i=1}^{N} \ln{(\sigma_{vr}^2+\sigma_i^2)}+\sum_{i=1}^{N} \frac{(v_i-< V_r >-\frac{\mathrm{d}v}{\mathrm{d}\chi}\chi_i)^2}{\sigma_{vr}^2+\sigma_i^2}\Bigg]
\end{equation}
where $v_i$ are the radial velocities of individual stars, $\sigma_i$ the errors on each star radial velocity and $\chi_i$ the angular distance between the photometric centre of the SMC and the $i$-th star projected along the gradient axis as follows:
\begin{equation}\label{gradient}
    \chi_i=(\alpha_i-\alpha_0)\mathrm{cos}(\delta_0)\mathrm{sin}(\theta)+(\delta_i-\delta_0)\mathrm{cos}(\theta)
\end{equation}
where ($\alpha_0,\delta_0$) are the photometric centre coordinates and ($\alpha_i,\delta_i$) are the coordinates of each individual star.
The results of this MCMC analysis were not constraining and we were not able to find the same velocity gradient that \citet{2014MNRAS.442.1663D} reported. 

We then implemented a Gaussian smoothing analysis to look for statistically significant spatial features. For this, we used a Gaussian kernel with a scale length equal to the projected distance to the $10$-th nearest neighbour (for each star) to smooth our extended data set. We then Monte Carlo sampled over the velocity errors, creating a Gaussian-smoothed map for each individual draw. From the median and $\pm68\%$ confidence intervals of these maps, we calculated a {\it smoothed} $V_r$ over the range: -6\fdg0 < $\alpha$ < 30\fdg0 and -77\fdg5 < $\delta$ < -68\fdg0, and a similarly smoothed distribution of uncertainties on $V_r$. We will refer to this as our \textquote{smoothed} data sample. We calculated the errors on the dispersion as in \citet{1993ASPC...50..357P} and \citet{2006ApJ...642L..41W}: 
\begin{equation}\label{dispersion}
    \sqrt{\sigma_{vr,i}^2}=\sqrt{\frac{\sum_{i=1}^{N}\rho_{i}V_{r,i}^2}{\sum_{i=1}^{N}\rho_i}-\Bigg[\frac{\sum_{i=1}^{N}\rho_{i}V_{r,i}}{\sum_{i=1}^{N}\rho_i}\Bigg]^2-\sigma_{i}^2}
\end{equation}
where $\sigma_{vr,i}$, $\rho_i$, $V_{r,i}$ and $\sigma_i$ are, respectively, the radial velocity dispersion, the density of the Gaussian kernel, the radial velocity and the error on the radial velocity of the $i$-th data point.
The errors, $\sigma_i$, for each data point were computed by Monte Carlo sampling  a Gaussian with a mean of zero and a width equal to the observational errors.

\begin{figure}
\includegraphics[width=\columnwidth]{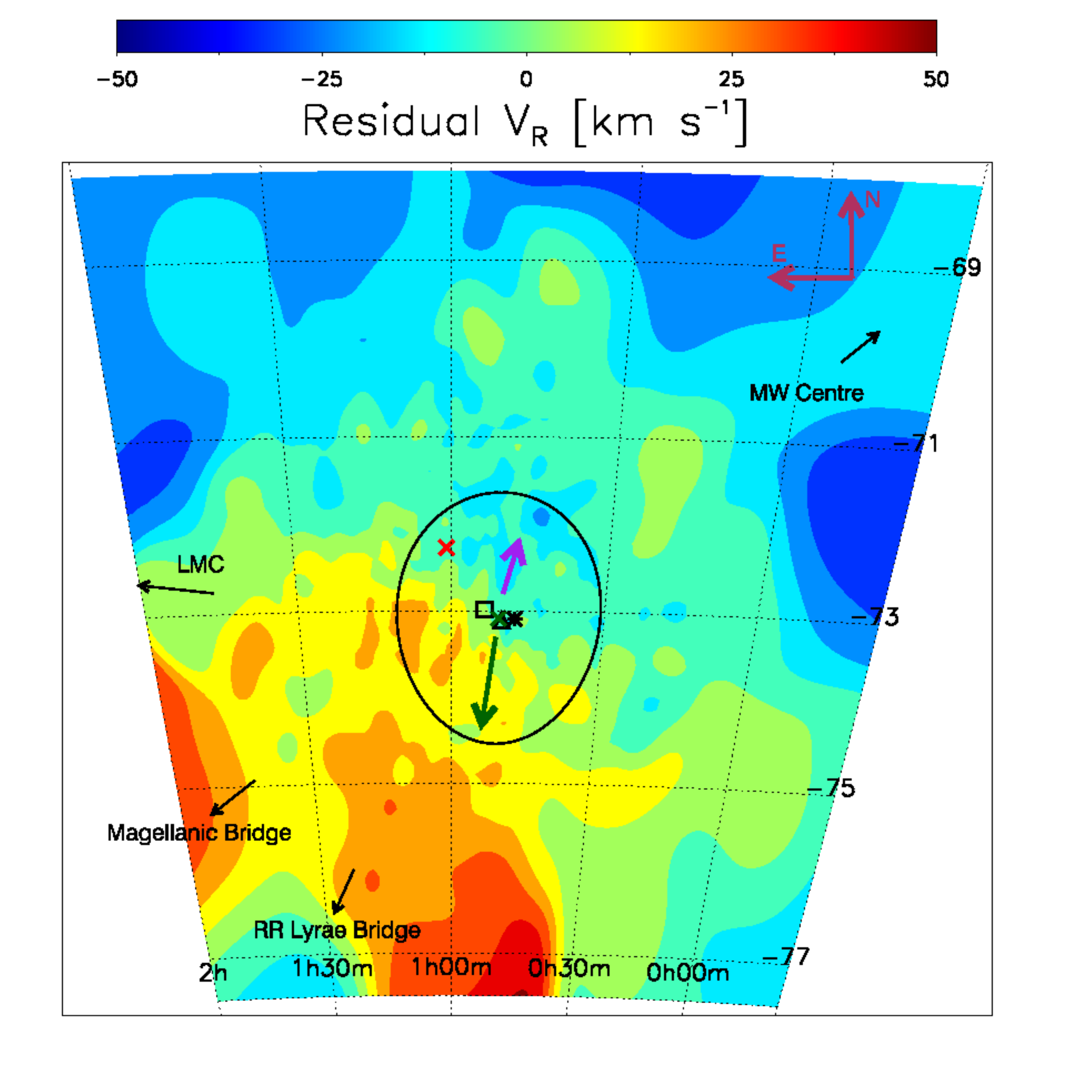}
\caption{Colour map in ($\alpha,\delta$) showing the residual radial velocity (after removing the SMC bulk motion) in km\,s$^{-1}$ of our smoothed sample of 34656 stars. The black arrows point in the directions of the LMC, MW centre, Magellanic Bridge (MB) and the RR Lyrae overdensity \citep{2017MNRAS.466.4711B}, the black ellipse represents the half-light radius at 1.5\,kpc, the dark green arrow represents the SMC mean proper motion, the purple arrow represents the SMC mean proper motion relative to the LMC. The green cross is the centre adopted for our analysis and the other symbols near it and overlapping each other (black asterisk, black square, back triangle) are, respectively, the maximum of the $\mbox{H\,{\sc i}}$ brightness \citep{1997MNRAS.289..225S}, the centre of the RR Lyrae distribution \citep{2012ApJ...744..128S} and the centre of the Classical Cepheids distribution \citep{2017MNRAS.472..808R}. The red cross is the $\mbox{H\,{\sc i}}$ kinematic centre identified by \citet{2019MNRAS.483..392D}. \label{fig:rv_map}}
\end{figure}

\begin{figure}
\includegraphics[width=\columnwidth]{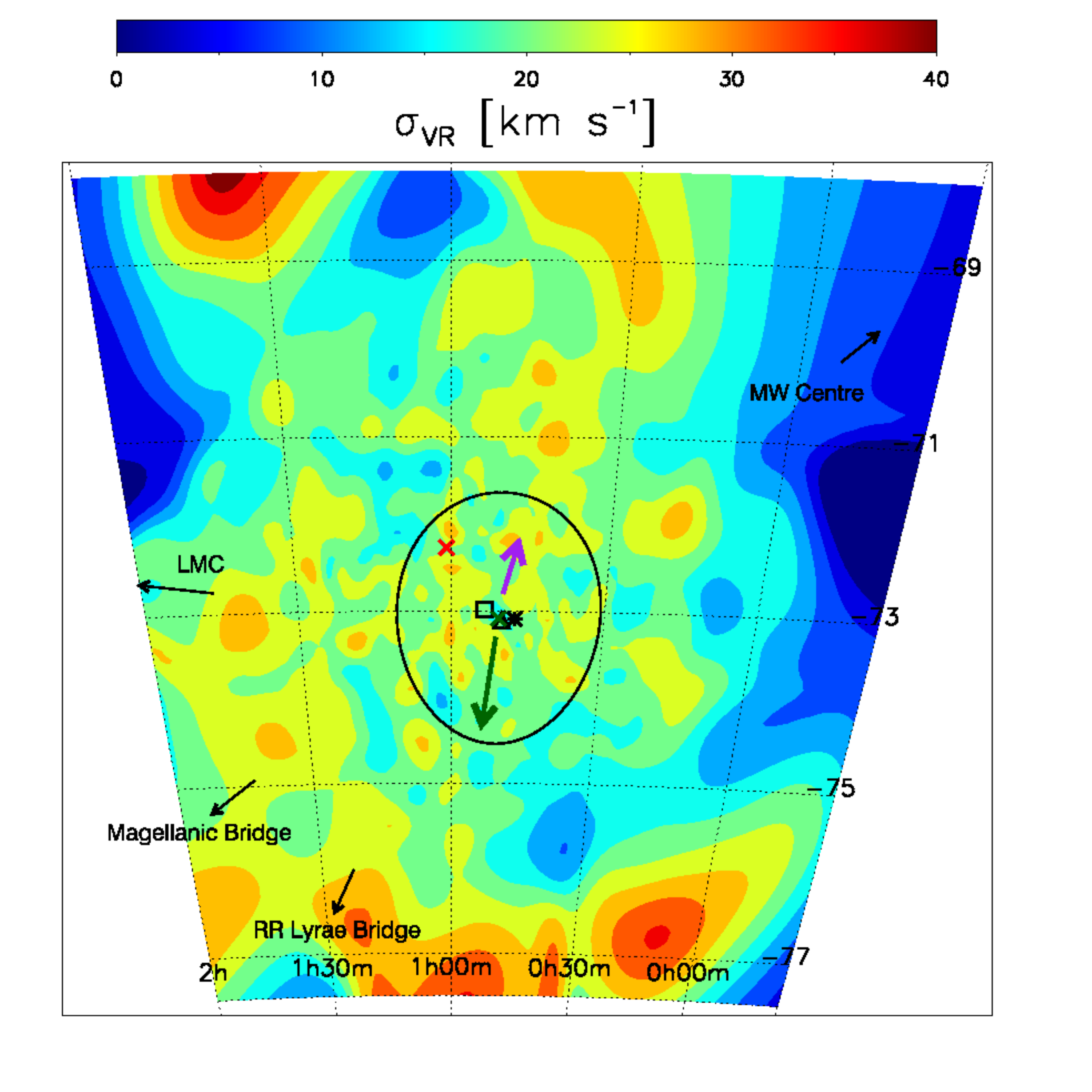}
\caption{Colour map in ($\alpha,\delta$) showing the radial velocity dispersion in km\,s$^{-1}$ of our smoothed sample of 34656 stars. The arrows and symbols are the same as in Fig.~\ref{fig:rv_map}. \label{fig:rvdisp_map}}
\end{figure}

The colour map in Fig.~\ref{fig:rv_map} shows the residual radial velocity (in km\,s$^{-1}$) for our smoothed sample. The colours from blue to red represent areas of lower to higher radial velocity with regards to the derived bulk motion of the SMC (as in Fig.~\ref{fig:spatial}, North, N, is up and East, E, is to the left of the reader). The black arrows point towards the various significant features and objects located outside of the area of the map, namely, the MW, the LMC,  the Magellanic Bridge and the RR Lyrae overdensity  (\citealt{2017MNRAS.466.4711B}; \citealt{2020ApJ...889...26J}). The solid black ellipse represents the half-light radius of 1.5\,kpc with the symbols inside the half-light radius denoting different estimations of the centre of the SMC (see caption of Fig.~\ref{fig:rv_map}). The green and purple arrow are, respectively, the SMC bulk motion relative to us and the SMC bulk motion relative to the LMC.

From Fig.~\ref{fig:rv_map} it is evident that there is a clear gradient inside the half-light radius, part of a larger-scale gradient (radial velocity increasing from NW to SE) that is split at the SE end with the two zones of higher radial velocity roughly in the directions of the Magellanic Bridge and the RR Lyrae overdensity.

The results from the radial velocity dispersion analysis are presented in Fig.~\ref{fig:rvdisp_map}. The colours from blue to red represent areas of, respectively, low and high velocity dispersion; the orientation and symbols are the same as those in Fig.~\ref{fig:rv_map}. This dispersion map shows slightly disturbed kinematics, even inside the half-light radius with dispersion values fluctuating around the mean value found by our maximum likelihood estimation (reported in Table~\ref{rv_smc_lit}). The map is mostly flat around $\sim$25 km\,s$^{-1}$ with some \textquote{bubbles} ($\sim$30 km\,s$^{-1}$) of high dispersion inside the half-light radius. We  tested if these \textquote{bubbles} were statistically significant by increasing the scale length of our Gaussian kernel (up to the distance to the $25$-th neighbour) to increase the smoothing and, separately, clipping our extended sample at 3-$\sigma$ and then 2-$\sigma$.
We also produced a map of the error on the estimation of the dispersion value, which was mostly flat around the value of 2 km\,s$^{-1}$, and subtracted it from the map of the dispersion. The latter test is an extreme case in which the statistical error is assumed to be always maximally overestimating the velocity dispersion, it is clearly a very conservative assumption but it serves the purpose of checking the statistical significance of the features in question.
The \textquote{bubbles} survived all of these conservative tests; a more detailed analysis to disentangle their physical meaning will be presented in a  future paper.
We omit the map of the errors on $V_r$ for brevity since we found these to be almost perfectly flat on the sky, with a mean error of 2.8\,km\,s$^{-1}$.

\section{\textit{Gaia} DR2 proper motions}\label{propmot}

\subsection{General considerations}\label{propmot_general}

The second data release of the \textit{Gaia} mission \citep{2018A&A...616A...1G} included proper motions and parallaxes for more than 1.3 billion sources with a limiting magnitude of $G$=21 mag. Cross-matching our extended sample with \textit{Gaia}~DR2 we found that 2585 of the 2588 the stars originally present in our own sample have proper motions and parallaxes in \textit{Gaia}~DR2, while in the case of the \citet{2014MNRAS.442.1663D} sample 4170 of the 4172 stars are in \textit{Gaia}~DR2. Restricting ourselves to the confirmed SMC members of our extended data set we end up with 5636 stars with radial velocity determinations and proper motions from \textit{Gaia}~DR2.

Due to the distance to the SMC ($\sim$60 kpc) and the faintness of the stars in our extended sample (G magnitude between 15 and 17 for 94\% of the sample), the parallaxes from the \textit{Gaia}~DR2 astrometric solution do not give a meaningful constraint on the distance (\citealt{2018A&A...616A..17A}; \citealt{2018A&A...616A...2L}; \citealt{2018A&A...616A...9L}). Since we already selected the SMC members with the radial velocity information, there is no need to rely on distance measurements to remove foreground MW stars. We checked the \textit{Gaia}~DR2 quality indicators for the stars in our sample and they all have 9 or more useful visibility periods, Astrometric Excess Noise (AEN) $< 2.1$ mas and astrometric sigma5d max $ < 0.6$ mas (as expected for stars with the full five parameter astrometric solution in the catalog, \citealt{2018A&A...616A...2L}). Throughout the analyses we will assume that all our stars are at the same distance.

Following the recommendations of L. Lindegren\footnote{IAU 30 GA \textit{Gaia} 2 astrometry talk, available in extended version at https://www.cosmos.esa.int/web/gaia/dr2-known-issues.}, we computed the total uncertainties of the \textit{Gaia} DR2 proper motion measurements using the following equation:

\begin{equation}
    \sigma_{tot}=\sqrt{k^2\sigma_i^2+\sigma_s^2}
\end{equation}

where $\sigma_i$ is the $i$-th measured uncertainty, $k=1.4$ is the correction factor to account for the underestimation of the measured uncertainties and $\sigma_s=0.066$ mas\,yr$^{-1}$ is the systematic uncertainty. The values of both $\sigma_s$ and $k$ are the estimates for stars in the magnitude range of our data sample (see \citealt{2018A&A...616A...2L} and the above mentioned talk by L. Lindegren). We will refer to the proper motion in right ascension, $\mu_{\alpha}\cos{\delta}$, simply as $\mu_{\alpha}$ (and variations of the symbol to indicate the mean proper motion or the one of individual stars).
For the proper motions analysis (Sec.~\ref{propmot_features}) we corrected the proper motion measurements to account for the solar reflex motion and then we added the further correction for geometric effects (systematic contraction/expansion with respect to the centre due to the motion along the line of sight) following \citet{2006A&A...445..513V} and \citet{2018MNRAS.481.2125B}:

\begin{equation}\label{geometric_eff}
    \mu_r=-6.1363\times10^{-5}V_{los}\frac{R}{d}
\end{equation}

where $\mu_r$ is the correction in mas\,yr$^{-1}$, $V_{los}$ is the velocity along the line of sightin km\,s$^{-1}$, $R$ is the distance to the centre of the SMC in arcmin and $d$ is the distance between the SMC and the Sun in kpc.

\subsection{SMC's bulk motion and proper motion distribution}\label{propmotbulk}

To characterise the proper motion distributions we used a MCMC maximum likelihood estimator as done for the radial velocity distribution. In this case we took into account the correlation between the $\mu_{\alpha}$ and $\mu_{\delta}$ and estimated the mean values for the proper motions (i.e. the bulk proper motion of the SMC) at the same time with the following likelihood:
\begin{equation}\label{likelihood_pms}
	\ln{\mathcal{L}}=-\frac{1}{2}\sum_{i=1}^{N}[(\mathbf{x_i}-\boldsymbol{\mu})^T\Sigma_i^{-1}(\mathbf{x_i}-\boldsymbol{\mu})+\ln{((2\pi)^2\det\Sigma_i)}]
\end{equation}
where $\mathbf{x_i}=(\mu_{\alpha,i}, \mu_{\delta,i})$ are the observed proper motion components,  $\boldsymbol{\mu}=(<\mu_{\alpha}>, <\mu_{\delta}>)$ are the mean values for the proper motions,  and $\Sigma_i$ is the correlation matrix (Eq.~\ref{corr_matrix}) of each star in the sample:
\begin{equation}\label{corr_matrix}
    \Sigma_i=\begin{pmatrix} \sigma_{\mu\alpha,i}^{*} & 0 \\ 0 & \sigma_{\mu\delta,i}^{*} \end{pmatrix}\begin{pmatrix} 1 & c_i \\ c_i & 1 \end{pmatrix}\begin{pmatrix} \sigma_{\mu\alpha,i}^{*} & 0 \\ 0 & \sigma_{\mu\delta,i}^{*} \end{pmatrix}
\end{equation}
where $\sigma_{\mu\alpha,i}^{*}$, $\sigma_{\mu\delta,i}^{*}$ and $c_i$ are, respectively, the observational errors on $\mu_\alpha$ and $\mu_\delta$ and their correlation coefficient, for each star. The intrinsic dispersion of the underlying distribution of the proper motions is not taken into account because, when factoring in the systematics explained in Section ~\ref{propmot_general}, the observational errors dominate the error budget. To prove this we tested the MCMC chain with a fixed $\sigma_{intr}=0.09$ mas\,yr$^{-1}$ (inferred from the radial velocity dispersion) for both $\alpha$ and $\delta$ and we recovered the same values for the mean proper motions of the SMC. Furthermore we ran the MCMC routine leaving the two intrinsic dispersions as indipendent free parameters and, despite recovering different values than the fixed one we used previously ($\sigma_{intr,\alpha}=0.194$ mas\,yr$^{-1}$ and $\sigma_{intr,\delta}=0.08$ mas\,yr$^{-1}$), the mean values of the proper motions changed from those reported in Table~\ref{pm_literature} by less than 0.01 mas\,yr$^{-1}$ (equivalent to less than 3 km\,s$^{-1}$), well within our uncertainties.

\begin{figure}
\includegraphics[width=\columnwidth]{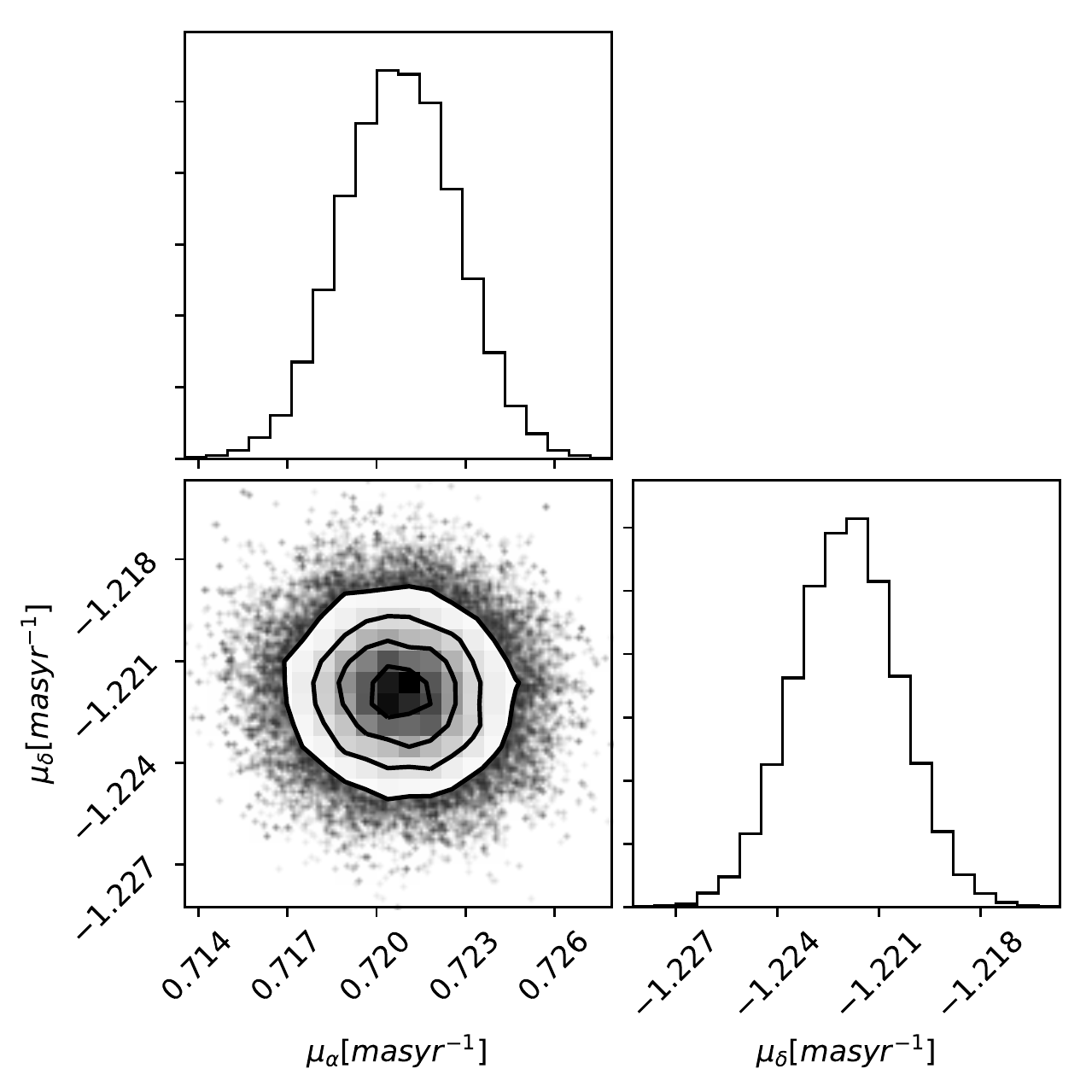}
\caption{Two dimensional and marginalised PDFs for the mean proper motions of the SMC, $<\mu_{\alpha}>$ and $<\mu_{\delta}>$. \label{fig:pms_mcmc}}
\end{figure}

The results of the MCMC can be seen in Fig.~\ref{fig:pms_mcmc} where we show the two dimensional Probability Density Function (PDF) for $<\mu_{\alpha}>$ and $<\mu_{\delta}>$ and their individual marginalised PDFs. The Pearson \textit{R} correlation coefficient for the proper motions is $\rho_R=-0.02$. Our final estimates of $<\mu_{\alpha}>$ and $<\mu_{\delta}>$ are reported in Table~\ref{pm_literature} as well as the values from \citet{2018ApJ...864...55Z} and \citet{2013ApJ...764..161K}. Our results are consistent with both of these studies, the first being one of the more recent estimates and the second being the generally accepted standard by the community. 

\begin{figure}
\includegraphics[width=\columnwidth]{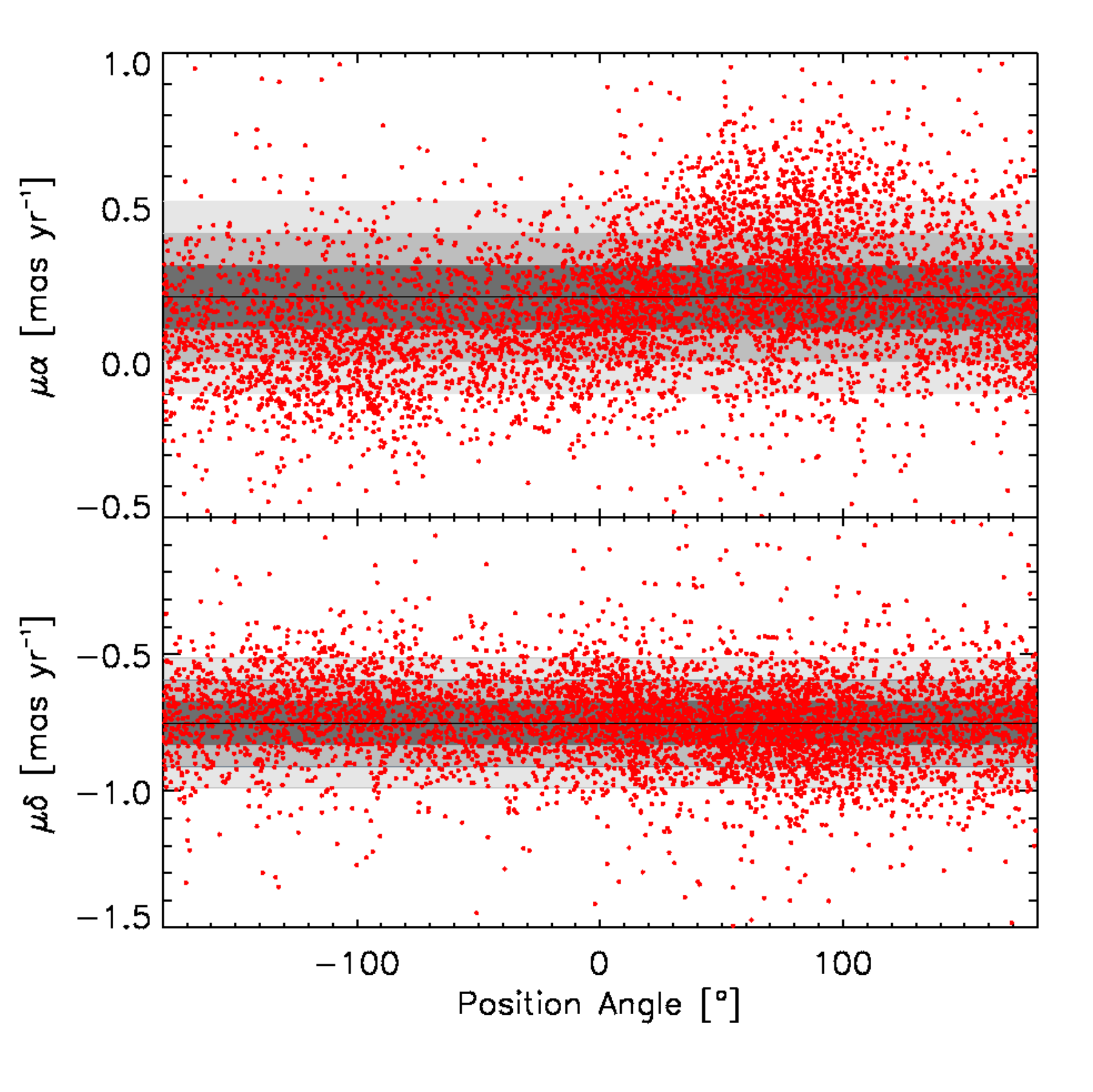}
\caption{Distribution of proper motions versus PA (defined positive North to East and negative North to West). {\it Top panel}: $\mu_{\alpha}$ versus PA, red points are individual stars, black line is the mean proper motion of the SMC, progressively lighter grey contours are the 1-$\sigma$, 2-$\sigma$ and 3-$\sigma$ intervals. {\it Bottom panel:} $\mu_{\delta}$ versus PA, symbols are the same as the top panel. \label{fig:pms_vs_pa}}
\end{figure}

Fig.~\ref{fig:pms_vs_pa} shows the distribution of the proper motions of the extended sample versus the PA defined positive North to East and negative North to West. The distribution of $\mu_{\alpha}$ is presented in the upper panel while the lower panel shows the distribution of $\mu_{\delta}$. Red dots are individual measurements, the black line is the mean value recovered with the MCMC and the progressively lighter grey contours are the 1-$\sigma$, 2-$\sigma$ and 3-$\sigma$ intervals. It is interesting to note the excess of stars with $\mu_{\alpha}$ more than 3-$\sigma$ higher than the mean value towards positive PA, in the direction of the LMC (PA $\sim$90$\degr$, East), and a similar excess of stars with $\mu_{\alpha}$ more than 3-$\sigma$ lower than the mean value,
in the opposite direction (PA $\sim$ -90$\degr$, West).

\begin{table}
 \centering
\caption{Proper motions for the SMC, the listed uncertainties include both statistical and systematic errors.\label{pm_literature}} 
 \begin{tabular}{@{}lcc@{}}
\hline
Study & $\mu_{\alpha}$ & $\mu_{\delta}$ \\
 & mas\,yr$^{-1}$ & mas\,yr$^{-1}$ \\
   \hline
Our data set $+$ \textit{Gaia}~DR2 & 0.690$\pm$0.220 & -1.180$\pm$0.127 \\
``Extended'' $+$ \textit{Gaia}~DR2 & 0.721$\pm$0.024 & -1.222$\pm$0.018 \\
\citet{2013ApJ...764..161K} & 0.772$\pm$0.063 & -1.117$\pm$0.061 \\
\citet{2018ApJ...864...55Z} & 0.82$\pm$0.12 & -1.21$\pm$0.04 \\
\hline
\end{tabular}
\end{table}

\subsection{Proper motions maps}\label{propmot_features}

To generate proper motion maps, we used the same kernel smoothing technique employed to generate our radial velocity smoothed sample (Sec.~\ref{RadVel_features}). 
To properly account for the correlation of the measurements in \textit{Gaia}~DR2, we performed the Monte Carlo sampling on a bivariate Gaussian distribution, extracting $\mu_{\alpha,i}$ and $\mu_{\delta,i}$ at the same time. The errors are likewise generated as in Sec.~\ref{RadVel_features} and the dispersions are computed through Eq.~\ref{dispersion} with the proper motions and their errors in place of the radial velocities and their errors.

\begin{figure}
\includegraphics[width=\columnwidth]{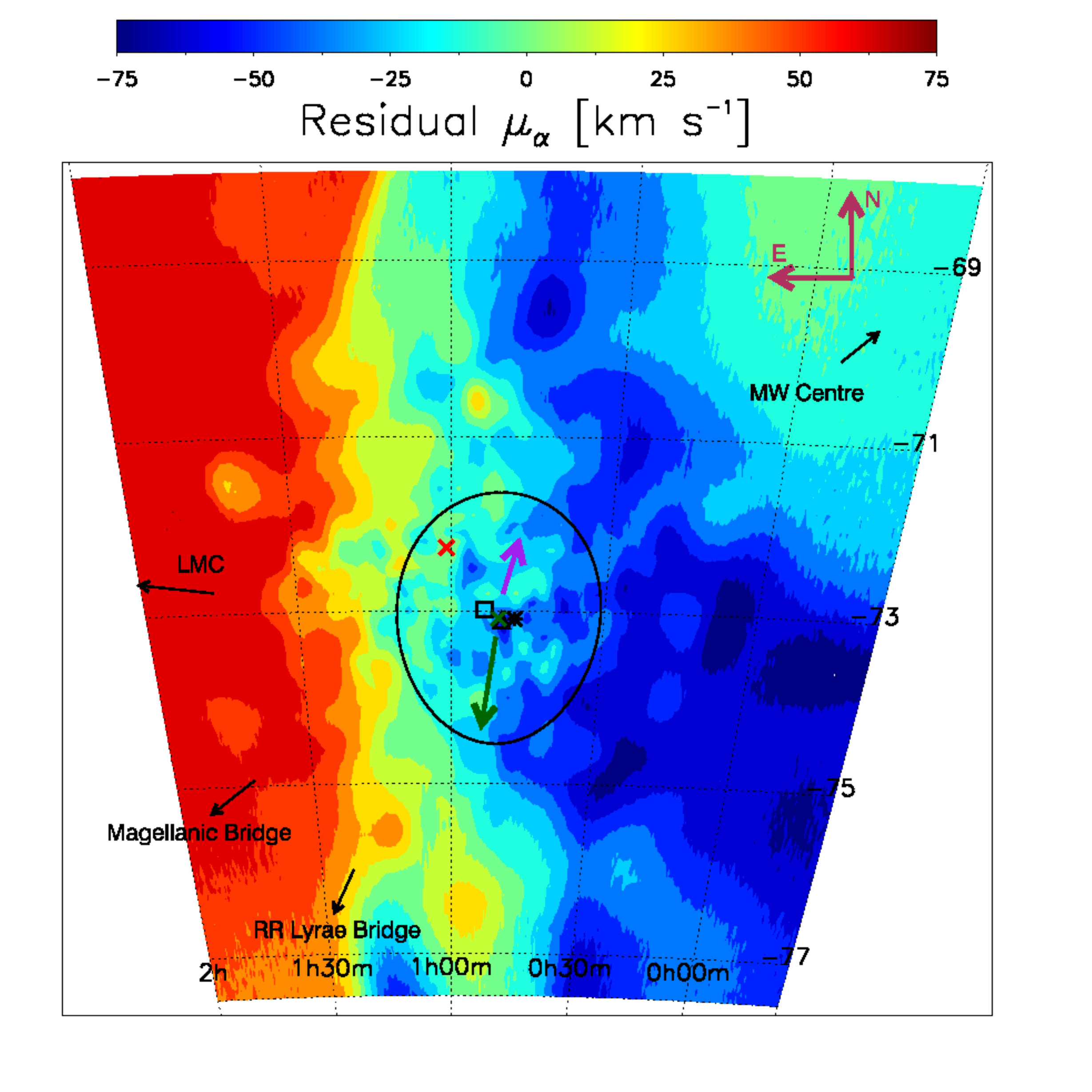}
\caption{Colour map in ($\alpha,\delta$) showing the residual proper motion $\mu_{\alpha}$ in km\,s$^{-1}$ of our smoothed sample of 34656 stars. The arrows and symbols are the same as in Fig.~\ref{fig:rv_map}. \label{fig:mura_map}}
\end{figure}

Fig.~\ref{fig:mura_map} shows the colour map of the spatial distribution of the residual $\mu_{\alpha}$ (after subtracting the SMC's bulk motion) converted into km\,s$^{-1}$ using the equivalence $v=4.74 \mu d$ (where the proper motion $\mu$ is in mas\,yr$^{-1}$ and the distance $d$ in kpc). The meaning of the arrows and symbols and the orientation of the map are the same as in Fig.~\ref{fig:rv_map}, the colours from blue to red show the areas of progressively higher residual velocity, positive in the direction of increasing $\alpha$ (towards East, to the left of the reader). There is a very conspicuous and clear gradient of increasing velocity going from West to East, in the direction of the LMC, unequivocally showing the SMC's disruption along this direction. 

\begin{figure}
\includegraphics[width=\columnwidth]{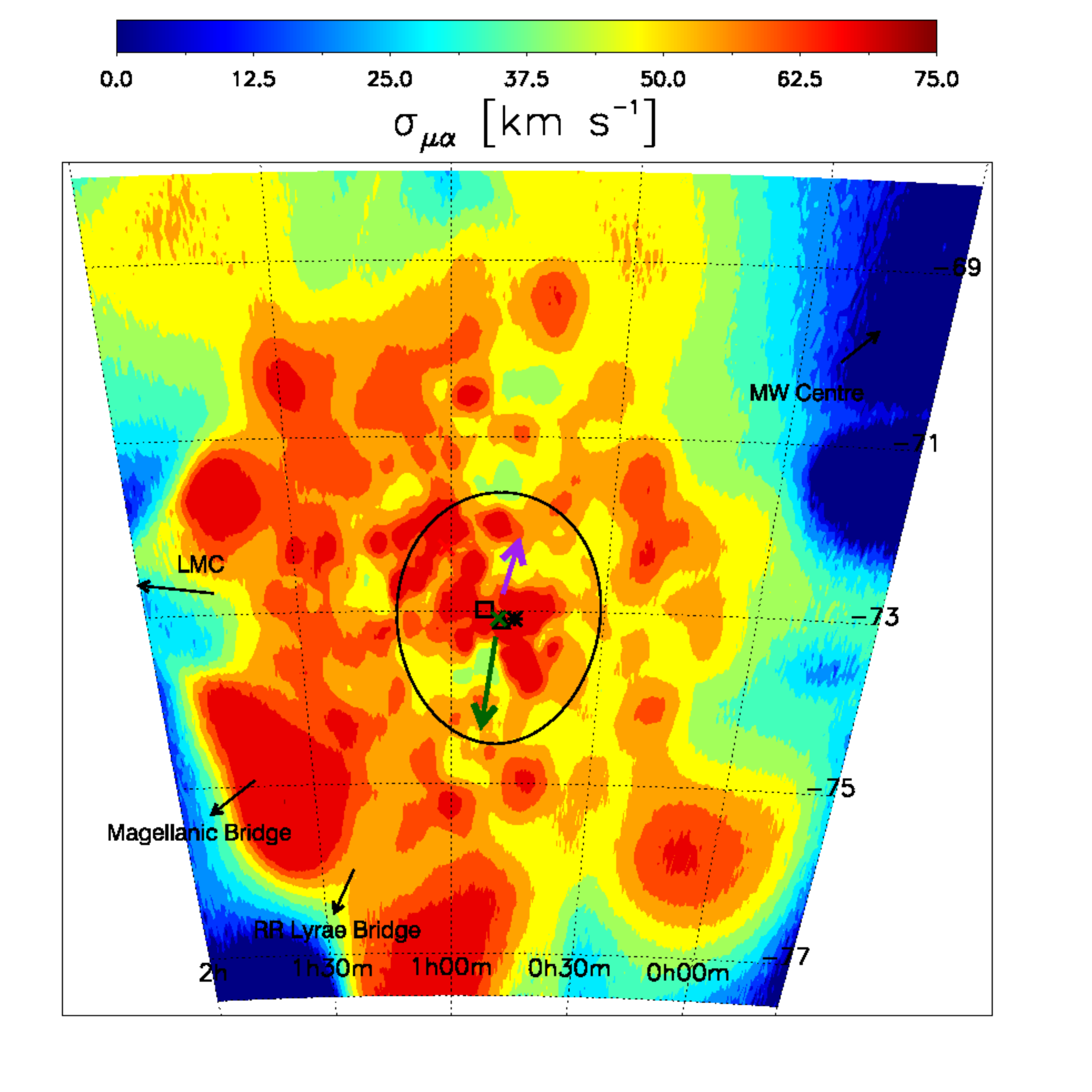}
\caption{Colour map in ($\alpha,\delta$) showing the dispersion $\sigma_{\mu\alpha}$ in km\,s$^{-1}$ of our smoothed sample of 34656 stars. The arrows and symbols are the same as in Fig.~\ref{fig:rv_map}. \label{fig:muradisp_map}}
\end{figure}

Fig.~\ref{fig:muradisp_map} shows the colour map of the spatial distribution of the dispersion in $\mu_{\alpha}$ converted to km\,s$^{-1}$ with symbols and colour conventions as in Fig.~\ref{fig:rv_map}. The map shows fluctuations around $\sim$50 km\,s$^{-1}$ and slightly disturbed kinematics inside the half-light radius with some high dispersion (>75 km\,s$^{-1}$) \textquote{bubbles}.
As in the case of the radial velocity maps, we omitted the error map in $\mu_{\alpha}$ because the errors are approximately flat around a mean error of 14.9\,km\,s$^{-1}$.

\begin{figure}
\includegraphics[width=\columnwidth]{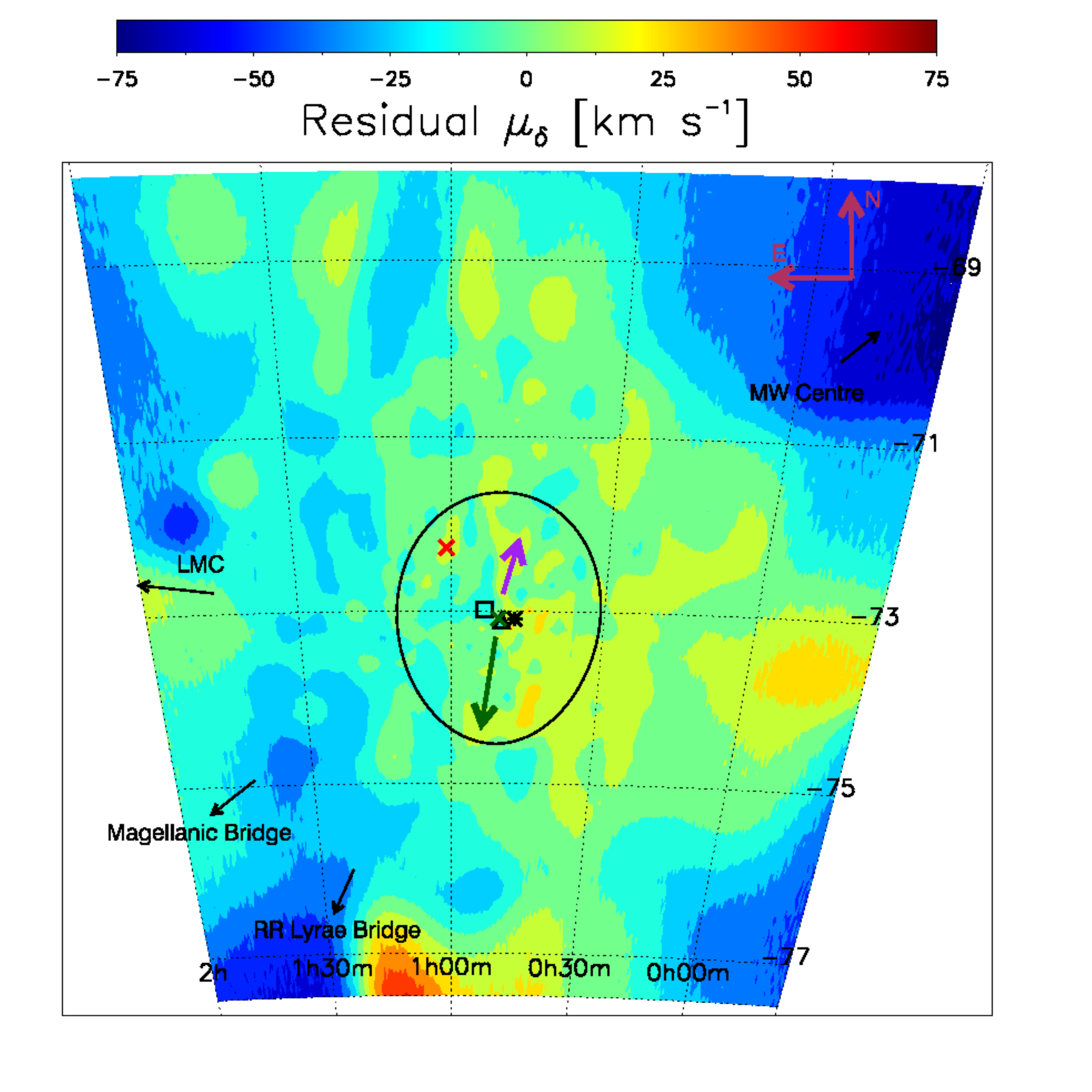}
\caption{Colour map in ($\alpha,\delta$) showing the residual proper motion $\mu_{\delta}$ in km\,s$^{-1}$ of our smoothed sample of 34656 stars. The arrows and symbols are the same as in Fig.~\ref{fig:rv_map}. \label{fig:mudec_map}}
\end{figure}

Fig.~\ref{fig:mudec_map} shows the colour map of the spatial distribution of the residual $\mu_{\delta}$ (after subtracting the SMC's bulk motion), converted into km\,s$^{-1}$. The meaning of the arrows and symbols and the orientation of the map are the same as in Fig.~\ref{fig:rv_map}. The colours from blue to red show areas of progressively higher residual velocity, where positive numbers are in the direction of increasing $\delta$ (towards North, to the top of the page). No significant gradient is found in this residual $\mu_{\delta}$ map.

\begin{figure}
\includegraphics[width=\columnwidth]{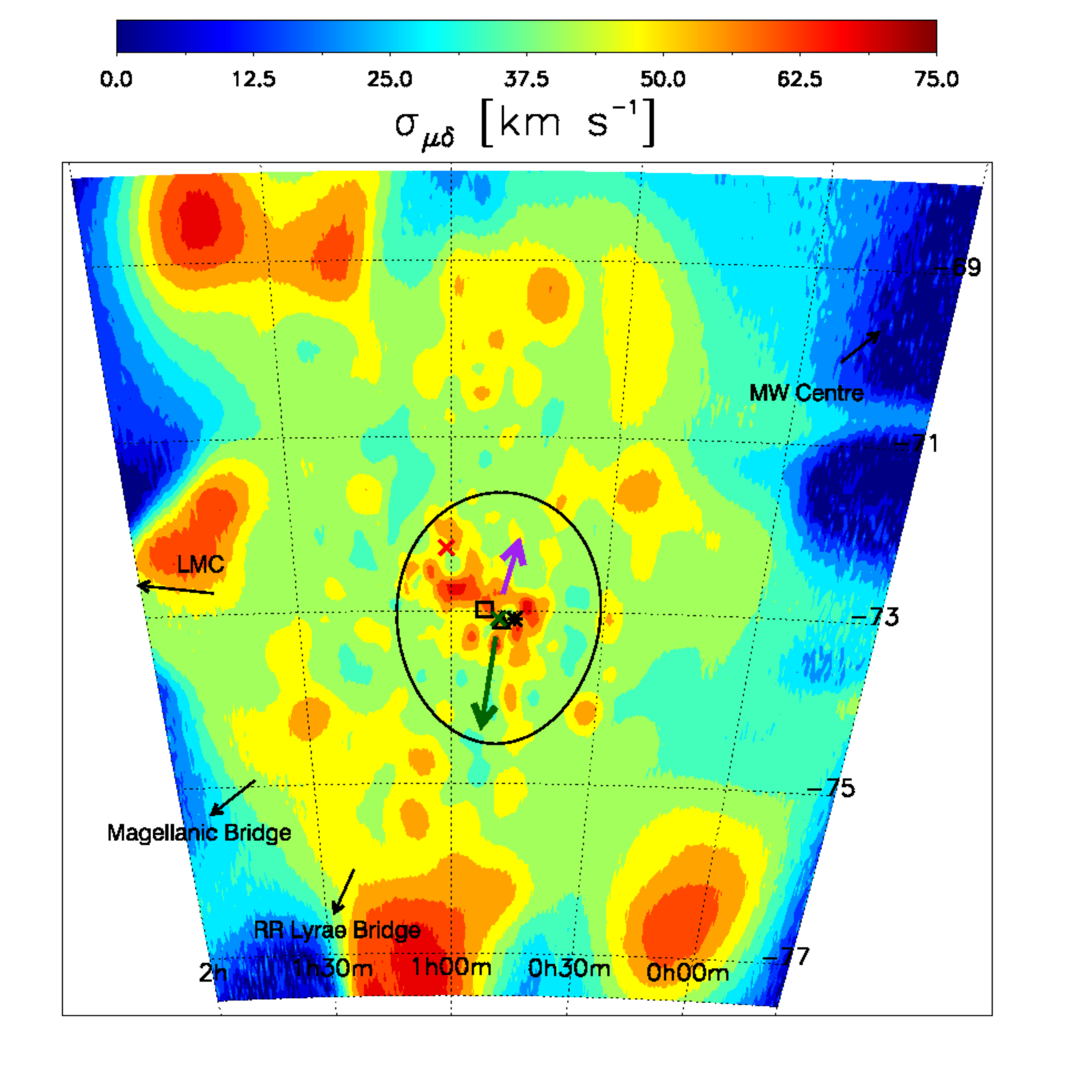}
\caption{Colour map in ($\alpha,\delta$) showing the dispersion $\sigma_{\mu\delta}$ in km\,s$^{-1}$ of our smoothed sample of 34656 stars. The arrows and symbols are the same as in Fig.~\ref{fig:rv_map}.\label{fig:mudecdisp_map}}
\end{figure}

Fig.~\ref{fig:mudecdisp_map} shows the colour map of the spatial distribution of the dispersion in $\mu_{\delta}$ converted to km\,s$^{-1}$. The meaning of the arrows and symbols and the orientation of the map are the same of Fig.~\ref{fig:rv_map}. The colours from blue to red show areas of progressively higher dispersion. As in Fig.~\ref{fig:muradisp_map}, despite being mostly flat around $\sim$40 km\,s$^{-1}$, this map shows slightly disturbed kinematics inside the half-light radius, with some high dispersion ($\sim$60 km\,s$^{-1}$) \textquote{bubbles}. As in the previous cases, the map of the errors in $\mu_{\delta}$ is omitted because the errors are flat around a mean error of 12.7\,km\,s$^{-1}$.

\begin{figure}
\includegraphics[width=\columnwidth]{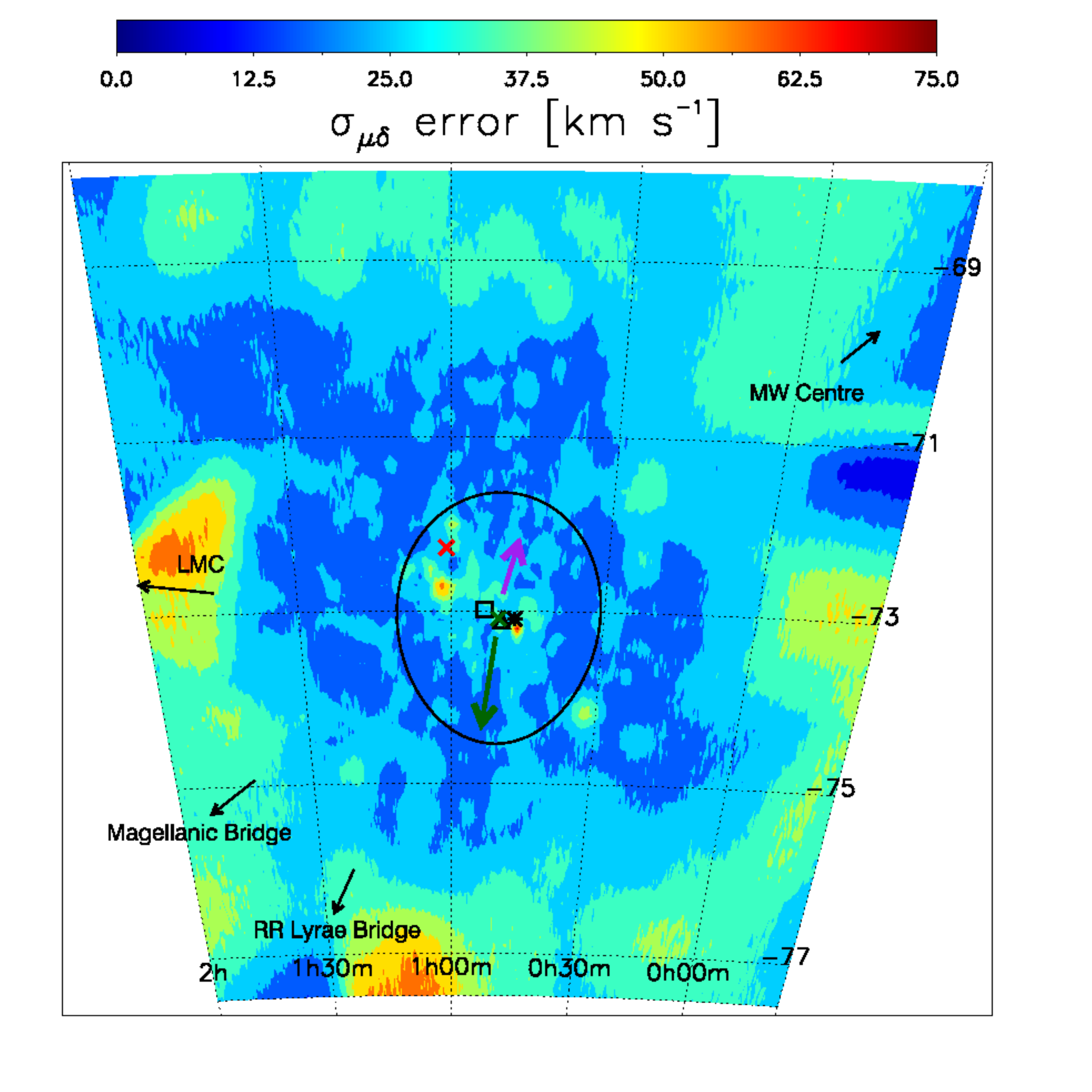}
\caption{Colour map in ($\alpha,\delta$) showing the error on the estimation of the dispersion $\sigma_{\mu\delta}$ in km\,s$^{-1}$ of our smoothed sample of 34656 stars. The arrows and symbols are the same as in Fig.~\ref{fig:rv_map}.\label{fig:mudecdisp_map_error}}
\end{figure}

\begin{figure}
\includegraphics[width=\columnwidth]{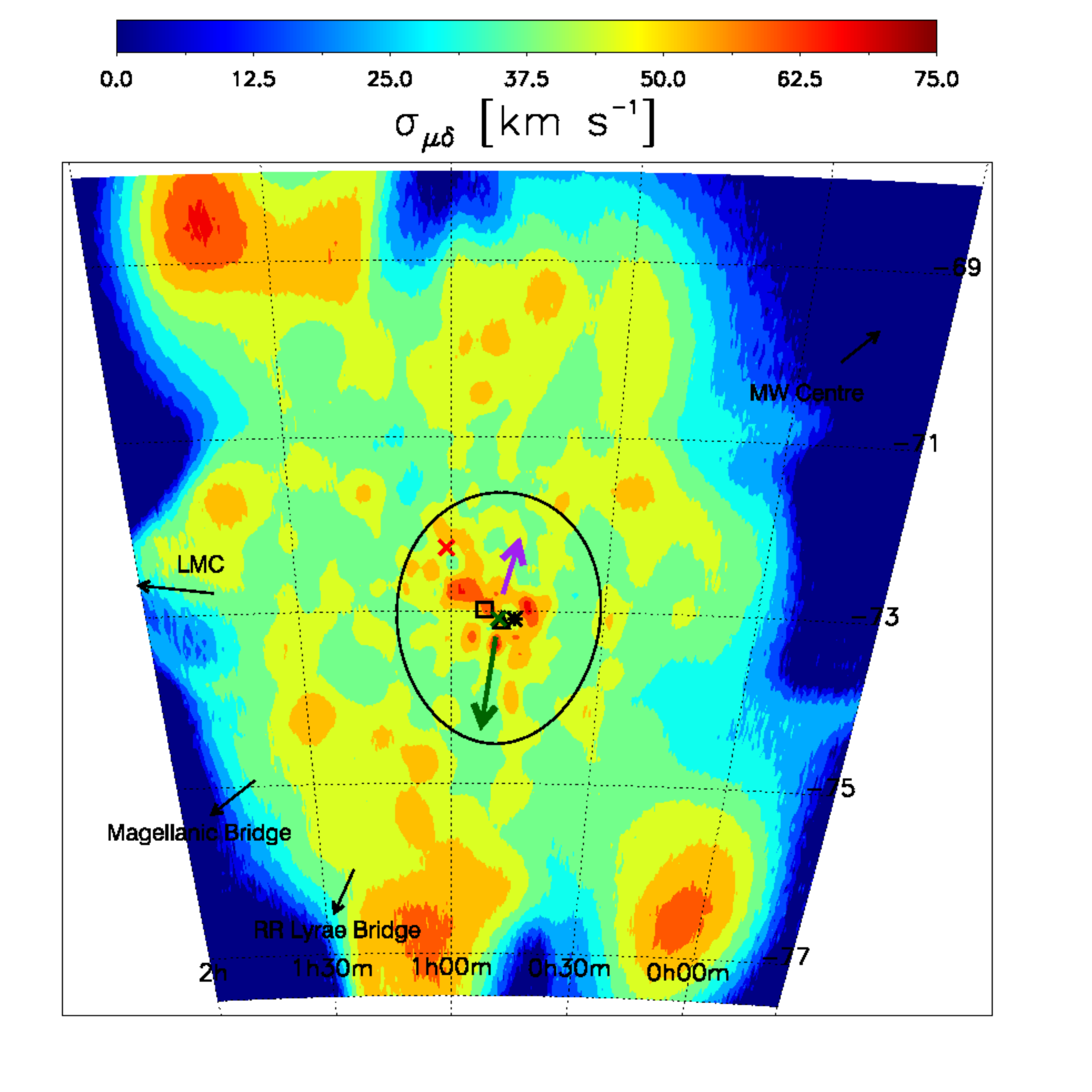}
\caption{Colour map in ($\alpha,\delta$) showing the dispersion $\sigma_{\mu\delta}$ in km\,s$^{-1}$ of our smoothed sample of 34656 stars after removal of the statistical error on its determination. The arrows and symbols are the same as in Fig.~\ref{fig:rv_map}.\label{fig:mudecdisperr_map}}
\end{figure}

We tested the robustness of the high dispersion features inside the half-light radius as we did for the radial velocity dispersion map. The tests with increased smoothing and $\sigma$-clipping the original data gave the same result (i.e. the \textquote{bubbles} did not disappear) while the maps of the errors in the estimation of the dispersions showed average errors in the footprint of our original data of 31.3\,km\,s$^{-1}$ for the errors in $\sigma_{\mu\alpha}$ and 25.9\,km\,s$^{-1}$ for the errors in $\sigma_{\mu\delta}$ and slightly larger errors outside of the footprint. Fig.~\ref{fig:mudecdisp_map_error} shows an example of a colour map in ($\alpha,\delta$) showing the errors on the estimation of the dispersion. 
 Despite the higher errors with respect to the radial velocity maps (which owes to the higher uncertainties on the original proper motion data), subtracting these new error maps from the dispersion maps did not erase the \textquote{bubbles}, as can be seen for example in Fig.~\ref{fig:mudecdisperr_map}. As stated before, we reserve a more in-depth analysis of the significance and physical meaning of these \textquote{bubbles} for future work.

\section{Discussion}\label{discussion}

\subsection{Tidal disruption of the SMC}\label{tidalSMC}

The proper motion analysis shows clear signatures of the tidal disruption of the SMC by the LMC. The stars in the Eastern side of the SMC's main body are being pulled faster towards the LMC than the stars in the main body and the stars located in the Western side are lagging behind, as it is apparent from Fig.~\ref{fig:mura_map}. No clear features are found in the orthogonal direction (Fig.~\ref{fig:mudec_map}). If this motion was caused by the ordered rotation of stars then we would expect that the features along the $\alpha$ direction would be maximised and increasing in magnitude outward along the North-South axis while features along the $\delta$ direction would be maximised and increasing in magnitude outward along the East-West axis. However, this is not what our analysis presented here shows; instead we find a complex picture, consistent with the tidal disruption of the SMC. This was already hinted at by \citet{2018ApJ...864...55Z}, briefly mentioned in \citet[][see the top left panel of their fig. 3 where the gradient in the proper motions is evident]{2019MNRAS.482L...9B} and implied by the smaller distance measured for the Eastern regions of the SMC \citep{2012ApJ...744..128S,2015MNRAS.449.2768D, 2017MNRAS.472..808R, 2018MNRAS.473.3131M} as they are being pulled towards the LMC and thus closer to us. This last point is corroborated by the gradient seen in Fig.~\ref{fig:rv_map} as the stars in the south-eastern part have higher radial velocities due to them being behind on the orbit of the SMC around the LMC (see the purple arrow in the figure for the direction of this orbit).
Coupled with the evidence of the stellar population in the Magellanic Bridge streaming towards the LMC found by \citet{2019ApJ...874...78Z}, our work shows that the kinematics of the SMC stellar population have been heavily disrupted by the interactions with the LMC.

\begin{figure*}
\includegraphics[width=\textwidth]{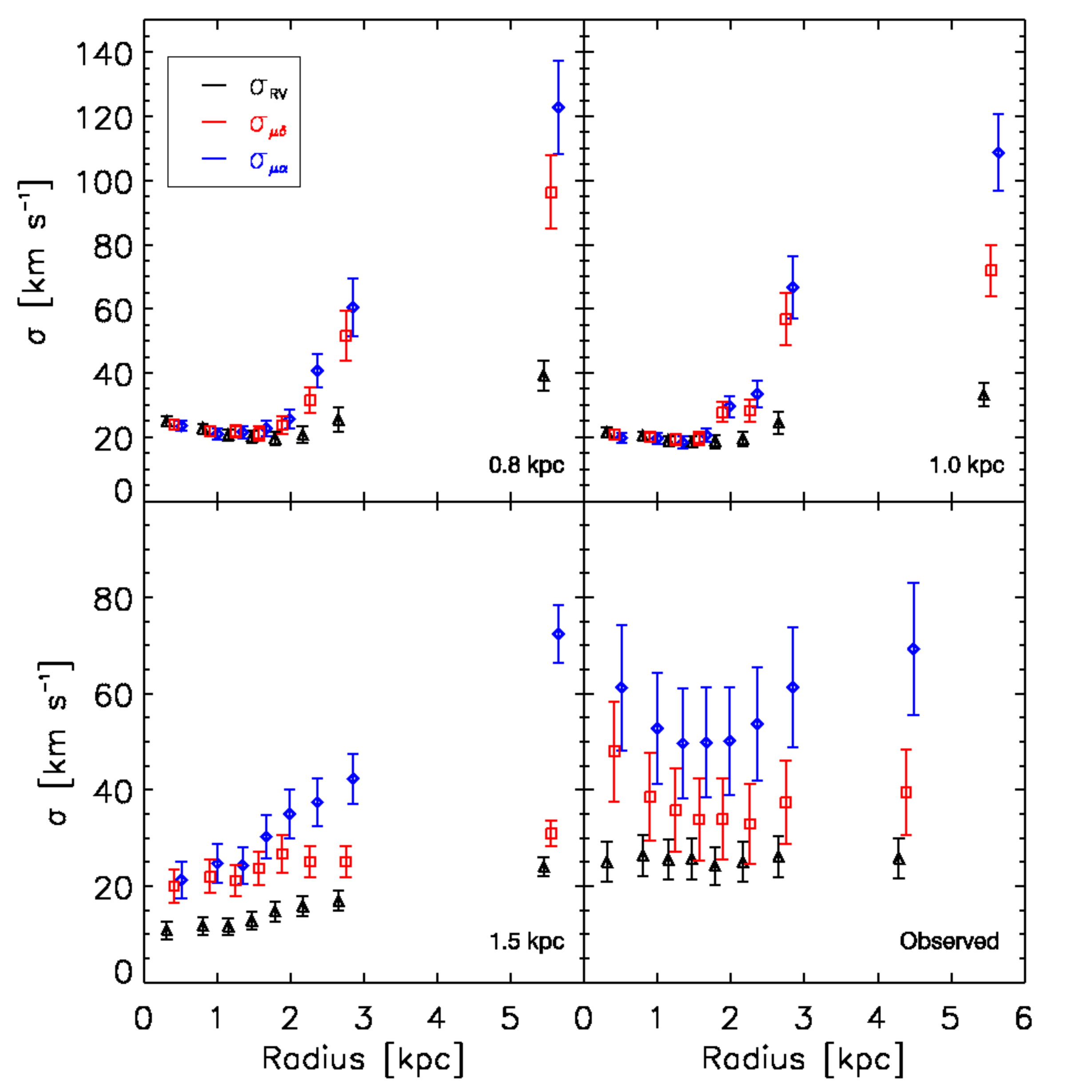}
\caption{Velocity dispersion profiles for the radial velocity (black), proper motion along $\delta$ (red) and proper motion along $\alpha$ (blue) binned in radius. The caption in the bottom right corner of each panel indicates the scale radius used for the SMC-like galaxy in the simulation (or, in the case of the final panel, for the observed data). In each panel, the data for the different dispersions have been computed in the same radial bins, they have then been artificially displaced along the X-axis for clarity. \textit{Top panels:} velocity dispersion profiles for simulations of tightly bound SMC-like galaxies. \textit{Bottom left panel:} velocity dispersion profile for a simulation of a disrupting SMC-like galaxy. \textit{Bottom right panel:} velocity dispersion profile derived from the observational data. Notice that as the simulated SMC scale length is increased from 0.8-1.5\,kpc (corresponding to an increasingly tidally disrupted SMC) the stars become increasingly tangentially anisotropic (the red and blue dispersion data are higher than the black), indicative of strong tidal stripping \citep{2007MNRAS.378..353K}. The real SMC data (bottom right panel) show similarly large tangential anisotropy, indicative of strong tides.\label{fig:disp_v_rad}}
\end{figure*}

To further explore the above scenario, we ran a grid of $N$-body simulations using \textsc{gadget-3} which is an updated version of \textsc{gadget-2} \citep{2005MNRAS.364.1105S}. The SMC was modelled as a Plummer sphere with a total mass (accounting for dark matter, gas and stars) of $M_{\rm SMC} = 10^9 M_{\odot}$ and a range of scale radii from 0.8 kpc to 1.5 kpc in steps of 0.1 kpc with $10^5$ particles. The LMC was modelled as a particle sourcing a Hernquist profile with a mass of $1.5\times10^{11} M_\odot$ and a scale radius of $17.14$ kpc, consistent with the recent measurement of the LMC's mass in \cite{2019MNRAS.487.2685E}. The MW was modelled using the \texttt{MWPotential2014} model from \cite{2015ApJS..216...29B}. We initialised the LMC and SMC using a mock realisation of their observed proper motions, radial velocities and distances. The pair was then rewound for 5 Gyr in the combined presence of each other and the MW. During this rewinding process, the SMC is modelled as a single particle of the above mentioned total mass sourcing its Plummer sphere potential. A live SMC modelled as a Plummer sphere is then injected and the SMC and LMC are evolved to the present.

The aim of this grid of simulations is to produce a suite of SMC-like galaxies that show similar features to the real SMC to form the basis of a qualitative comparison with our observational data. The simulation process produces an LMC/SMC-like pair which is roughly, but not exactly, at the present day position of the actual galaxies. This is because during the rewinding procedure the SMC-like galaxy is treated as a rigid Plummer sphere with a fixed mass while during the forwards integration it is allowed to disrupt. We note that the initial mass used for the SMC is based on its present-day dynamical mass \citep[e.g.][]{2006AJ....131.2514H} which is substantially less than its peak mass \citep[e.g. $\sim 5\times10^{10} M_\odot$ in][]{2019MNRAS.487.5799R}. This difference in mass is motivated by the fact that the simulations in this work are not meant to faithfully model the entire disruption history of the SMC but to rather model the final phases of this once the SMC has lost the majority of its dark matter \citep{2016ApJ...833..109S}.

From our grid of simulations, we extract three models of an SMC-like galaxy, each with a different scale radius for the density profile. We focus on the velocity dispersion profiles of the stars along the line of sight, and in the $\alpha$ and $\delta$ directions on the sky, binned in angular distance from the photometric centre. To ensure that we make a fair comparison, we use the same analysis pipeline for both observational and simulated data, adding to the latter mock-observational errors by sampling from a Gaussian distribution with zero mean and a dispersion equal to the mean observational error on the relevant variable (e.g. the radial velocity). The simulations produced data across the whole sky but we selected only the star particles with coordinates inside our observational window (the same used for the smoothed data sample): -6\fdg0 < $\alpha$ < 30\fdg0 and -77\fdg5 < $\delta$ < -68\fdg0.

The results of our comparisons are shown in Fig.~\ref{fig:disp_v_rad}. The black, red and blue symbols with errorbars show the dispersion in radial velocity, proper motion in the $\delta$ direction and proper motion in the $\alpha$ direction, respectively (as indicated by the legend in the top left panel). The scale radii used for the SMC-like galaxy in each specific simulation are also included as legends in the figures. All velocity dispersion profiles for all panels (simulations and observations) are computed in the same radial bins (the displacement along the X-axis of the data points in Fig.~\ref{fig:disp_v_rad} has been introduced for clarity). The only exception to this is the last radial bin, which is different between simulations and observational data because, in the same patch of sky, we had particle data in the simulations at angular distances slightly higher than the maximum angular distance of our observed sample, so the simulation bin extends slightly further.

The simulated SMC-like galaxies with tightly bound stars (i.e. with small scale radii; top panels of Fig.~\ref{fig:disp_v_rad}) are more difficult to tidally disrupt. Nonetheless, a clear kinematic signature of tidal disruption is seen at large radii in this simulation in the form of tangential velocity anisotropy. This can occur as stars moving on prograde and radial orbits are more easily stripped than those on retrograde orbits \citep[e.g.][]{2006MNRAS.366..429R}, or it can occur due to contamination from unbound stars being projected along the line of sight \citep[e.g.][]{2007MNRAS.378..353K}. In our simulations, the latter effect dominates (compare the dispersion profiles in Figures \ref{fig:disp_v_rad} and \ref{fig:disp_v_rad3kpc}). Simulated SMC-like galaxies with stars that are less bound (i.e. with larger scale radius; bottom left panel of Fig.~\ref{fig:disp_v_rad}) show the same tangential velocity anisotropy feature, but at smaller radii. In the least bound simulation with a scale radius of $1.5$\,kpc, this feature extends all the way to the centre of the SMC, similarly to the real data. This provides further evidence that the SMC is in an advanced stage of tidal disruption.

Although the simulations do not exactly reproduce the observed SMC (in particular, the real SMC has systematically higher dispersion and is likely, therefore, more massive than the simulated SMCs), they provide a clear indication that the tangential anisotropy seen in the data (bottom right panel of Fig.~\ref{fig:disp_v_rad}) is due to tidal effects induced by the LMC.

\subsection{Centres of the SMC}\label{centreSMC}

The tidal disruption of the SMC is the main kinematic signature emerging from our data. We find no signs of ordered rotation in the RGB stellar population, nor in the proper motion data. The absence of a detection of rotational signal in the stellar motions does not necessarily imply that the SMC RGB population does not have a level of ordered rotation but that the hypothetical rotation is dominated by the kinematical signature of the tidal effects. While it is feasible for a galaxy as massive as the SMC to have had a fully formed stellar disc at some point in its past \citep{2016MNRAS.459...44T}, such a disc is not detectable now through kinematic studies due to the dominant effect of the tidal interactions with the LMC.
In the presence of strong tidal effects on the stellar kinematics, it is surprising that the $\mbox{H\,{\sc i}}$ gas can retain an ordered rotating component as claimed by \citet{2019MNRAS.483..392D}. In fact, there are hints of massive gas outflows being stripped from the SMC \citep{2018NatAs...2..901M} and the gas mass budget of the Magellanic system shows that the Clouds have lost a considerable amount of gas to their surrounding features due to their tidal interactions. The gas in the Magellanic Stream, Magellanic Bridge and Leading Arm is twice as much as the combined gas content of the MCs ( see \citet{2005A&A...432...45B} and \citet{2014ApJ...787..147F}), with most of the gas in the Magellanic Stream probably being from the SMC. \citet{2017ApJ...836L..13K} show that a combination of tides, ram pressure stripping and feedback (from SNe and other processes) can transform rotating dwarf irregular galaxies into very slowly rotating dwarf spheroidal galaxies in a few Gyrs. The presence of gas, that we know the SMC had in the past, accelerates this process. The above mentioned evidence that most of the gas of the SMC has been evacuated to the surrounding features testifies that the SMC is partway through this process of tidal stirring. 
The implication of this body of evidence is that it is not possible anymore to extract a clear rotational signal from the motion of stars and gas in the SMC.
This in turn means that the $\mbox{H\,{\sc i}}$ kinematic centre, which is significantly displaced from both the $\mbox{H\,{\sc i}}$ `maximum brightness' centre \citep{1997MNRAS.289..225S} and numerous photometric centres -- old stellar populations \citep{2009A&A...496..375G}, RR Lyrae \citep{2012ApJ...744..128S} and Classical Cepheids \citep{2017MNRAS.472..808R}, all in good agreement with one another -- is not indicative of the true centre of the galaxy.  Recent work by \citet{2019ApJ...887..267M} shows that the rotating disc model for the gas cannot reproduce the observed motions of young O and B stars, typically used as gas tracers.
It is possible, therefore, that the velocity gradient observed in the $\mbox{H\,{\sc i}}$ data owes to ram pressure stripping or outflows, rather than ordered rotation. We will consider this idea in more detail in future work.

\subsection{Structure of the SMC}\label{structureSMC}

The available distance determinations for individual objects inside the SMC imply a large spread of line of sight depth for the SMC depending on the tracers used and  region that is being observed. Most observations point at a line of sight depth above 5\,kpc with extremes of even 20\,kpc \citep[see][for a review]{2015AJ....149..179D}. 
This observed great depth of the SMC along line of sight implies an extended bound structure (as concluded by \citealt{2017MNRAS.472..808R} and \citealt{2018MNRAS.473.3131M}) that is at odds with the highly perturbed kinematics we found here and the signs of interaction within  $\sim2-2.5$\,kpc from the SMC centre found by \citet{2017MNRAS.467.2980S} in the Magellanic Bridge.

\begin{figure}
\includegraphics[width=\columnwidth]{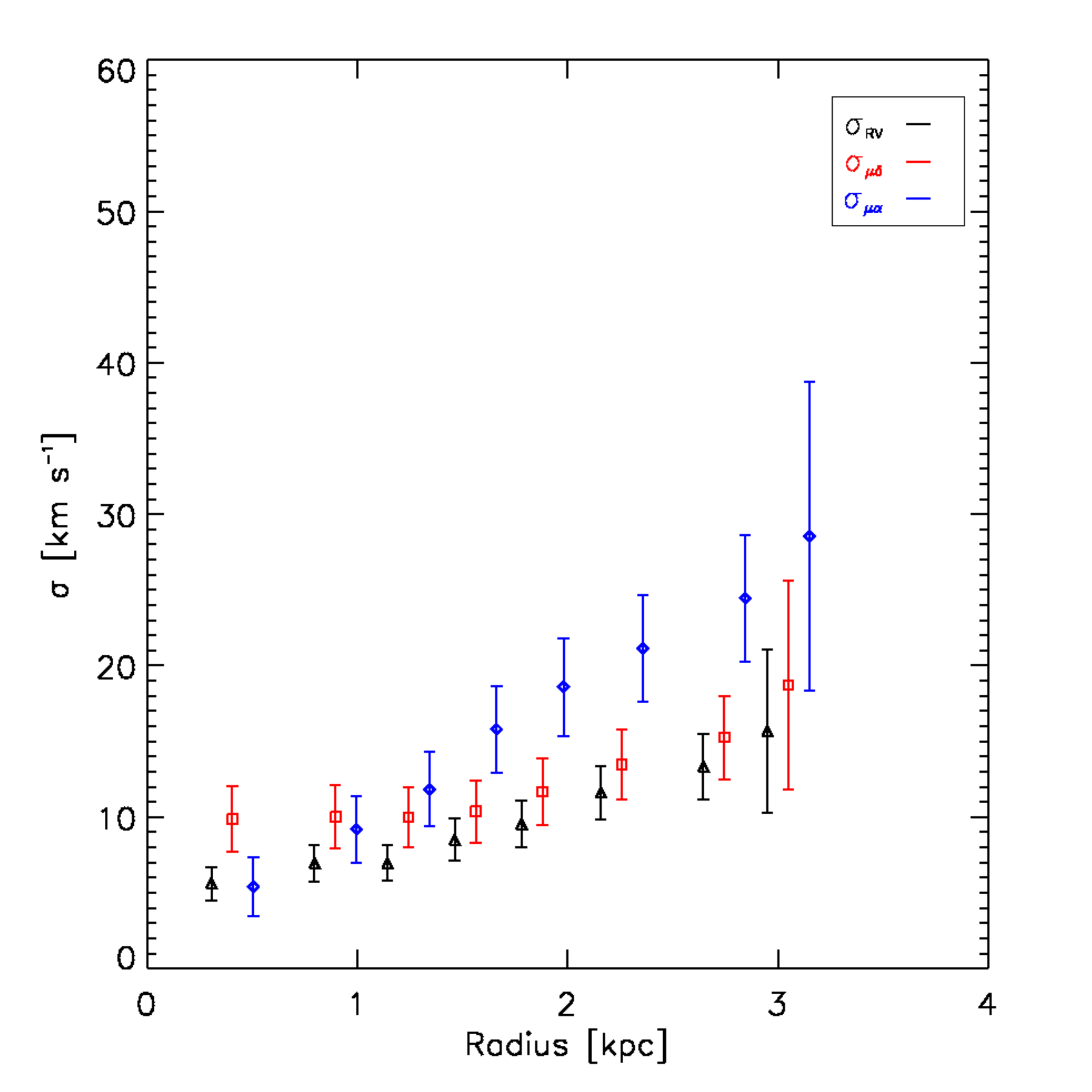}
\caption{Velocity dispersion profiles binned in radius of the stars within 3 kpc of the centre of the SMC-like galaxy in our most disrupted simulation (1.5 kpc scale radius). The symbols and legend are the same as in Fig.~\ref{fig:disp_v_rad3kpc}.\label{fig:disp_v_rad3kpc}}
\end{figure}

The strong evidence of tidal effects disrupting the SMC that we found and the measurements of great line of sight depth can be reconciled if most of the surviving bound structure of the SMC resides within the inner $\sim$2\,kpc while the rest of the objects along the observed depth are made of tidally stripped debris preceding or following the SMC core in its orbit. If this is the case then the high dispersion \textquote{bubbles} that we see are probably due to the tidal debris along the line of sight that would have different motions than the stars in the SMC's main body \citep[as shown for dwarf galaxies in general by][]{2007MNRAS.378..353K}.

Supporting evidence for this interpretation can be gleaned by our most disrupted simulation where if we restrict to a sphere of radius 3 kpc around the centre we find only a third of the particles in it and the kinematics inside the half-light radius become much more relaxed with no tangential anisotropy within 1.5 kpc (see Fig.~\ref{fig:disp_v_rad3kpc}). This implies that the disturbed kinematics (the \textquote{bubbles}) within the half-light radius might be due in large part to a projection effect by the debris along the line of sight and when these are removed the kinematics appear much more relaxed. It is worth pointing out that Fig.~\ref{fig:disp_v_rad3kpc} does not disprove our claim of heavy tidal disruption since the kinematics are clearly perturbed down to around 1.5 kpc even for this much refined sample.  In future work, we will perform a more thorough analysis aimed at disentangling the effects of projected tidal debris and tidal effects on the kinematics.

\section{Conclusions}\label{conclusions}

We analysed new spectroscopic data for $\sim$3000 SMC RGB stars observed with the AAOmega spectrograph at the Anglo-Australian Telescope, existing spectroscopic measurements from \citet{2014MNRAS.442.1663D}, and proper motions from the \textit{Gaia} DR2 catalogue in order to study the kinematics, internal structure, and dynamics of the SMC. The picture of the SMC that emerges from our analysis is that of a galaxy that is undergoing heavy tidal disruption by the LMC, enough to leave a clear kinematic signature on the old RGB population, as evidenced by a net outward motion of stars in the direction of the LMC and apparent tangential anisotropy not only in the outskirts but also all the way into the very centre of the SMC. Using a suite of $N$-body simulations of the MCs orbiting the MW, we showed that this tangential anisotropy likely owes to material unbound from the SMC, that has distinct kinematics, being projected along the line of sight.

Finally, we found a clear and conspicuous offset between the photometric and gas kinematics centres of the SMC. In particular, the advanced state of tidal disruption we highlight as the main kinematic signature of the RGB population of the SMC calls into question assumptions of steady state equilibrium for both gaseous and stellar kinematic dynamical models. The rotating disc model, on the basis of which the gas kinematic centre was identified, appears to be insufficient to explain the complex kinematics of the system. We will consider this in more detail in future work.

\section*{Acknowledgements}

The authors thank the anonymous referee for their enlightening comments. We would also like to thank  Sne\v{z}ana Stanimirovi{\'c} for providing access to the $\mbox{H\,{\sc i}}$ data. MDL would like to thank Davide Massari and Alexandre Vazdekis for helpful discussions.
The research leading to these results has received funding from the European Community's Seventh Framework Programme (FP7/2013-2016) under grant agreement number 312430 (OPTICON).
CG acknowledges financial support through the grants (AEI/FEDER, UE) AYA2017-89076-P, as well as by the Ministerio de Ciencia, Innovaci{\'o}n y Universidades (MCIU), through the State Budget and by the Consejer{\'i}a de Econom{\'i}a, Industria, Comercio y Conocimiento of the Canary Islands Autonomous Community, through the Regional Budget.
This work has made use of data from the European Space Agency (ESA) mission
{\it Gaia} (\url{https://www.cosmos.esa.int/gaia}), processed by the {\it Gaia}
Data Processing and Analysis Consortium (DPAC,
\url{https://www.cosmos.esa.int/web/gaia/dpac/consortium}). Funding for the DPAC
has been provided by national institutions, in particular the institutions
participating in the {\it Gaia} Multilateral Agreement. This research made use of  Astropy \citep{2013A&A...558A..33A, 2018AJ....156..123A}, Matplotlib \citep{2007CSE.....9...90H} and Numpy \citep{A guide to NumPy, 2011CSE....13b..22V}.

\appendix

\section{Observational Data Table}

This appendix contains the first 10 rows of our table of observational data, the complete form of which is available as supplementary material online.

\begin{table*}
 \centering
\caption{Observational data, columns are as follow: 2MASS identifier, RA coordinate, Dec coordinate, radial velocity determined as specified in Section ~\ref{RadVelsDet}, error on radial velocity determination and signal-to-noise ratio. (This table is availbale in it's entirety online in machine-readable form)  \label{data_table}} 
 \begin{tabular}{@{}lccccc@{}}
\hline
ID & RA [$\degr$] & Dec [$\degr$] & RV [km\,s$^{-1}$] & RVerror [km\,s$^{-1}$] & SNR \\
   \hline
2MASS00031888-7230310 & 0.82867 & -72.50863 & 115.83 & 0.64 & 12.0 \\
2MASS00033588-7310330 & 0.8995 & -73.17586 & 133.5 & 0.92 & 13.2 \\
2MASS00051882-7258454 & 1.32846 & -72.9793 & 130.99 & 0.29 & 24.3 \\
2MASS00053603-7309222 & 1.40017 & -73.15618 & 139.29 & 0.2 & 31.2 \\
2MASS00053827-7302468 & 1.40946 & -73.04634 & 170.76 & 0.31 & 26.6 \\
2MASS00061378-7240169 & 1.55742 & -72.67138 & 159.77 & 0.41 & 22.0 \\
2MASS00063269-7310123 & 1.63621 & -73.1701 & 138.9 & 0.45 & 23.9 \\
2MASS00074727-7256140 & 1.947 & -72.93725 & 131.73 & 0.52 & 18.4 \\
2MASS00083937-7220479 & 2.16408 & -72.34664 & 141.57 & 0.44 & 20.2 \\
2MASS00084051-7253142 & 2.16879 & -72.88728 & 137.98 & 0.68 & 17.0 \\
\hline
\end{tabular}
\end{table*}

\bsp
\label{lastpage}
\end{document}